\documentclass[10pt,journal]{IEEEtran}
\IEEEoverridecommandlockouts
\ifCLASSINFOpdf

\else

\fi

\usepackage[cmex10]{amsmath}
\usepackage{amssymb}
\usepackage{cite}
\usepackage{graphicx}
\usepackage{array,color}
\usepackage{amsmath}
\usepackage{stfloats}
\usepackage{graphicx}
\usepackage{subfigure}
\usepackage{tabularx}
\usepackage{epsfig,epsf,color,balance,cite}
\usepackage{algorithmicx}
\usepackage{algorithm}
\usepackage{bm}
\usepackage{textcomp}
\usepackage[justification=centering]{caption}
\usepackage{cuted}
\usepackage{makecell}
\usepackage{subcaption}
\newtheorem{remark}{Remark}

\begin{document}
	\title{Modular XL-Array-Enabled 3-D Localization\\ based on Hybrid Spherical-Planar Wave Model\\ in Terahertz Systems}
	\author{Yang Zhang, Ruidong Li, Cunhua Pan, \emph{Senior Member, IEEE}, Hong Ren, \emph{Member, IEEE}, \\Tuo Wu, Changhong Wang

		\thanks{(Corresponding author: \emph{Cunhua Pan and Hong Ren})
			
			Yang Zhang, Cunhua Pan, and Hong Ren are with the National Mobile Communications Research Laboratory, Southeast University, Nanjing 210096, China (E-mail: \{220230982, cpan,  hren\}@seu.edu.cn).
			
			R. Li and C. Wang are with Shandong Yunhai Guochuang Cloud Computing Equipment Industry Innovation Co., Ltd., Jinan, China (E-mail: \{lird, wangchh01\}@inspur.com).
			
			T. Wu is with the School of Electrical and Electronic Engineering, Nanyang Technological University, 639798, Singapore (E-mail:
			tuo.wu@ntu.edu.sg).
			}
	}

	\markboth{MANUSCRIPT SUBMITTED TO IEEE Transactions on Cognitive Communications and Networking }%
	{XL Modular Array-Enabled 3-D Localization based on Hybrid Spherical-Planar Wave Model in Terahertz Systems}
	
	\maketitle
	\vspace{-1.5cm}

\begin{abstract}
This work considers the three-dimensional (3-D) positioning problem in a Terahertz (THz) system enabled by a modular extra-large (XL) array with sub-connected architecture.
Our purpose is to estimate the Cartesian Coordinates of multiple user equipments (UEs) with the received signal of the RF chains while considering the spatial non-stationarity (SNS).
We apply the hybrid spherical-planar wave model (HSPWM) as the channel model owing to the structual feature of the modular array, and propose a 3-D localization algorithm with relatively high accuracy and low complexity.
Specifically, we first distinguish the visible sub-arrays (SAs) located in the VR and estimate the angles-of-arrival (AoAs) from each UE to typical visible SAs with the largest receive power via compressed sensing (CS) method.
In addition, we apply the weighted least square (WLS) method to obtain a coarse 3-D position estimation of each UE according to the AoA estimations.
Then, we estimate the AoAs of the other SAs with a reduced dictionary (RD)-CS-based method for lower computational complexity, and utilize all the efficient AoA estimations to derive a fine position estimation.
Simulation results indicate that the proposed positioning framework based on modular XL-array can achieve satisfactory accuracy with evident reduction in complexity.
Furthermore, the deployment of SAs and the allocation of antenna elements need to be specially designed for better positioning performance.
\end{abstract}

\IEEEpeerreviewmaketitle

\section{Introduction}\label{intrdc}
\IEEEPARstart{A}{ccurate} spatial positioning information has become not merely an auxiliary function but a foundational capability that enables critical applications in the sixth generation (6G) communications, such as autonomous driving, industrial automation, extended reality (XR), and ultra-precise healthcare monitoring \cite{zhou2020service, zhang2019wireless, hu2020cellular}.
These applications entail high localization accuracy, and promote wireless positioning networks as a promising technique\cite{liu2007survey}.
In general, wireless positioning systems aim at estimating channel parameters, such as angles of arrival (AoAs) and delays, thereby obtaining the position estimation of user equipment (UE).
In recent years, the millimeter-wave (mmW) is considered as the essential frequency band for accurate channel parameter estimation and localization\cite{dardari2015indoor, lin2018indoor, han2020millimeter, wu2022ris, wu2023fingerprint, wu2024joint, wu2024exploit}.
Nevertheless, the accuracy of mmW-based algorithms is still constrained by limited bandwidth, which calls for further development in 6G wireless positioning network.

Fortunately, the Terahertz (THz) band, typically ranging from 0.1 THz to 10 THz, is considered as critical 6G waveband owing to abundant spectrum resources and support of extra-large multiple-input-multiple-output (XL-MIMO)\cite{cacciapuoti2018beyond, rappaport2019wireless, chen2019survey}.
Specifically, the tiny wavelength enables the collaboration of hundreds of antenna elements (AEs) on one array, constituting the framework of XL-MIMO, which is capable of combating the severe path loss in THz band with high array gain.
Hence, the THz band is highly anticipated to fulfill the positioning accuracy demand with high signal quality and resolution, showing the potential of achieving subcentimeter-level accuracy in localization \cite{xing2021millimeter, kanhere2021position}, and offering unprecedented opportunities to achieve the extreme positioning requirements of future intelligent systems.

Nevertheless, owing to the combined impact of tiny wavelength and relatively large array aperture, the currently applied planar wave model (PWM) for the far field (FF) may no longer be valid and should be substituted by spherical wave model (SWM) in the near field (NF) \cite{selvan2017fraunhofer, cui2022channel, han2023toward, wang2024tutorial}.
In other words, the phenomenon of cross-field communication becomes not negligible in XL-MIMO enabled THz systems, and traditional localization algorithms cannot be directly applied in these scenarios.
Recently, the study on NF localization algorithm has been flourishing \cite{abu2021near, elzanaty2021ris, rahal2021ris, rinchi2022compressive, he2021mixed, he2022mixed, pan2023ris, huang2024low, lu2024near}.
Notably, the authors of \cite{pan2023ris} proposed a near-field joint channel estimation and localization (NF-JCEL) algorithm in THz systems with the assistance of a reconfigurable intelligent surface (RIS).
Based on the proposed model, they demonstrate that more RIS elements functioning as localization anchors can provide more angular information, thus leading to higher angular resolution when estimating the AoAs in NF scenarios.
In addition to this, the authors of \cite{huang2024low} simplify the model and proposed a low-complexity NF-JCEL algorithm for XL-MIMO-enabled THz systems.

However, implementing accurate and efficient localization in THz XL-MIMO systems presents several fundamental challenges.
In XL-array-enabled systems, the large aperture results in the problem of spatial non-stationarity (SNS), where some AEs may be blocked from receiving signals of certain UEs. This is owing to obstacles or the inherent directionality of THz signals, and thereby fundamentally alters the channel characteristics\cite{yuan2023spatial}.
Meanwhile, the existing SWM-based NF localization algorithms involve high-dimensional operations on large covariance matrices, resulting in prohibitive computational demands that scale unfavorably with array size.
In addition, traditional XL-MIMO implementations also require dedicated radio frequency (RF) chains for each AE, leading to excessive hardware costs, power consumption, and implementation complexity which limit practical deployment.
Furthermore, achieving both high-precision positioning and computational efficiency in XL-MIMO-enabled scenarios creates a fundamental tension between using complex SWM (accurate but computationally intensive) and simplified PWM (efficient but potentially inaccurate).

Fortunately, the modular XL-array architecture \cite{li2024multiuser,li2024near} has recently been proposed as a promising approach for THz communications. This architecture consists of multiple sub-arrays (SAs) sharing the same configuration, regularly deployed on a planar surface with inter-SA spacing significantly larger than the spacing between adjacent AEs. This modular design offers several significant advantages: Firstly, it effectively addresses the hardware complexity issues in traditional XL-MIMO systems by decomposing a single giant antenna array into multiple smaller SAs, each requiring only one RF chain, substantially reducing hardware costs and power consumption\cite{yu2016alternating, ayach2014spatially, liu2024hybrid, chuang2015high, shu2018low, zhang2022doa}. Secondly, the modular architecture enhances system scalability and flexibility, allowing for adaptable adjustment of SA quantities and layouts according to application requirements. Thirdly, the increased inter-SA spacing creates a larger effective aperture, significantly improving angular resolution and positioning accuracy potential, particularly for distant targets. Fourthly, the modular design enhances system robustness, as the system can continue to function even if some SAs experience interference or failure. Finally, from an engineering implementation perspective, standardized SA production and assembly greatly simplify the manufacturing process, which significantly improves cost-effectiveness. These advantages collectively make the modular XL-array an ideal choice for high-precision positioning in the THz band.

While the employment of modular XL-MIMO offers these substantial benefits, its application to 3-D localization introduces new technical challenges. Firstly, with limited RF chains in hybrid beamforming (HBF) architectures, determining which SAs provide the most valuable positioning information becomes critical, particularly when the problem of SNS affects signal reception. Then, simultaneously achieving both high accuracy and low complexity requires the development of a multiple-stage approach which progressively refines the position estimates while preventing error propagation between stages. Finally, the distributed nature of SAs creates a complex multi-dimensional parameter space that requires specialized estimation algorithms capable of handling diverse angular information from spatially separated sources.

Particularly, to deal with the last challenge, it is reasonable to adapt the hybrid spherical-planar wave model (HSPWM), which balances the spherical wave characteristics between SAs with planar wave assumptions within each SA. This model has been applied to specific topics in terms of modular XL-array, such as cross-field channel estimation\cite{tarboush2024crossfield} and beam alignment\cite{chen2024can}. Recently, some of the studies also considered the problem of NF user positioning based on the HSPWM channel model in XL-array-enabled systems \cite{qiao2024sensing, li2024machine}. Nonetheless, these positioning methods utilized geometric relationship only to obtain the final position estimation, while it also can be reconsidered as an effective property to reduce the complexity in the stage of channel parameter estimation. In addition, the angular information obtained from different SAs are not weighted during the process of integration to form reliable position estimates, which accounts for varying estimation qualities across SAs.



To address these complex challenges, we propose a novel three-stage 3-D positioning framework specifically designed for modular XL-MIMO-enabled THz systems. Our approach intelligently combines power-based SA selection, CS-based angle estimation, and geometrically-guided dictionary reduction to achieve both high  positioning accuracy and computational efficiency.

The main contributions of this paper are summarized as follows: 

\begin{itemize}
	\item[1)]
	We consider the 3-D positioning problem in a sub-connected modular XL-array-enabled THz system.
	According to the structural feature, we propose a 3-D localization algorithm composed of three stages to obtain the coordinate estimation of the UEs, where each stage is strategically designed to address specific challenges.
	
\end{itemize}

\begin{itemize}
	\item[2)]
	In \textit{Stage 1}, we tackle the visible SA selection problem by identifying the VR corresponding to each UE and selecting typical visible SAs based on the received signal power, effectively managing the SNS issues.
	We then apply the simultaneous orthogonal matching pursuit (SOMP) method to obtain high-quality AoA estimations for these selected SAs. 
\end{itemize}

\begin{itemize}
	\item[3)]
	In \textit{Stage 2}, we address the challenge of integrating diverse angular information by developing a weighted least squares (WLS) method that accounts for estimation errors, producing a coarse but robust position estimation that serves as a foundation for subsequent refinement.
\end{itemize}

\begin{itemize}
	\item[4)]
	In \textit{Stage 3}, we resolve the computational complexity challenge through an innovative reduced dictionary (RD)-based method that leverages the coarse position estimates to dramatically shrink the search space for AoA estimation of remaining SAs. This stage-wise approach progressively refines the position estimate while maintaining computational efficiency, effectively balancing the trade-off between the accuracy of spherical wave models and the efficiency of planar wave approximations in our hybrid model.
\end{itemize}

\begin{itemize}
	\item[5)]
	Simulation results indicate that the  modular-array-enabled framework outperforms its collaborated counterpart in accuracy and robustness. In addition, careful design of SA interval and AE allocation can contribute to better positioning performance.
\end{itemize}

The remainder of this paper is organized as follows. The system model is introduced in Section II. The stages of the localization algorithm are proposed in Section III to V. Simulation results are provided in Section VI, and Section VII briefly concludes the paper.

\emph{Notations}: Constants, vectors and matrices are denoted by italics, boldface lowercase and boldface uppercase letters, respectively. ${{\mathbb{C}}^{M \times N }}$ denotes the set of $M \times N$ complex vectors or matrices. ${{\mathbb{E}}}\{\cdot\} $ denotes the expectation operation. ${\left\| {\bf{x}} \right\|}_{2}$ denotes the 2-norm of vector ${\bf{x}}$. ${\left\| {\bf{A}} \right\|_{\rm{F}}}$ and ${\rm{tr}}\left( {\bf{A}} \right)$ denote the Frobenius norm and trace of ${\bf{A}}$, respectively. ${ \bf{A} ^{*}}$, ${ \bf{A}^{\rm{T}}}$ and ${  \bf{A} ^{\rm{H}}}$ denote the conjugate, transpose and Hermitian transpose of $\bf{A}$, respectively. ${\bf{A}}_{[m,n]}$ denotes the $\left ( m, n \right ) $-th entry of ${\bf{A}}$. ${\rm{diag}}(\cdot)$ represents the diagonalization operator.  ${\bf{B}} \otimes {\bf{C}}$ and ${\bf{B}} \odot {\bf{C}}$ denote the Kronecker product and Hadamard product of ${\bf{B}}$ and ${\bf{C}}$, respectively. $\partial f ( x  ) /\partial x$ denotes the first-order partial derivative of function $f$ w.r.t.variable $x$. ${\bf{I}}_N$ denotes the $N \times N$ identity matrix, and ${\bf{r}} \sim {\cal C}{\cal N}({\bf{0}},{\bf{I}})$ denotes a random vector following the Gaussian distribution of zero mean and unit variance.

\section{System Model}\label{sysmod}
We consider the uplink transmission of a THz TDD 3-D localization system as shown in Fig. 1, where the BS equipped a with modular XL-MIMO array receives the pilot signals from $P$ single-antenna UEs for the positioning procedure.
The THz-band system has $I$ subcarriers, where the bandwidth and the center frequency are respectively denoted by $B$ and $f_c$, and the frequency of the $i$-th subcarrier is given as $f_i = f_c + \frac{B}{I}\left( i-\frac{I-1}{2}\right), i = 1, 2,\cdots, I$. Furthermore, we assume the existence of potential obstructions and scatterers in the considered scenario, which contributes to the SNS property of the XL-MIMO channel.

\begin{figure}[htbp]
	\centering
	\includegraphics[width=1.8in]{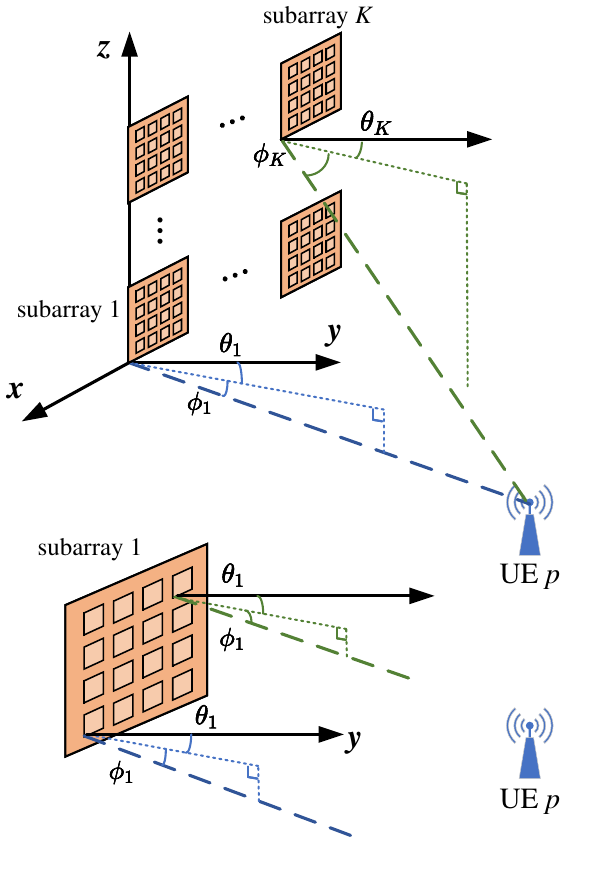}
	\caption{System Model.}
	\label{fig1}
\end{figure}
\subsection{Modular Architecture and Transmission Model}\label{asamod}
For simplicity in geometric relation and further manipulations, we consider that the modular XL-MIMO array is deployed parallel to the $xOz$ plane.
In addition, the $M$ antennas are uniformly divided into $K$ SAs, each SA is set as a UPA with $M_x \times M_z = M_{\rm{S}}$ antennas, and the $K$ SAs are deployed in a $K_x \times K_z$ uniform planar form with horizontal and elevation interval $D$.
Within each SA, the spacing between adjacent antennas is $d = \lambda_c/2$, where $\lambda_c = c/f_c$ denotes the wavelength corresponding to the central frequency.
Hence, by setting the location of the first SA as $(0, 0, 0)$, the location of the $k$-th SA is given as ${\bf{q}}^{\rm{B}}_k = \left(-(k_x-1)D, 0, (k_z-1)D\right)$, where $k = k_x + K_x(k_z - 1)$.

\begin{figure}[htbp]
	\centering
	\includegraphics[width=2.6in]{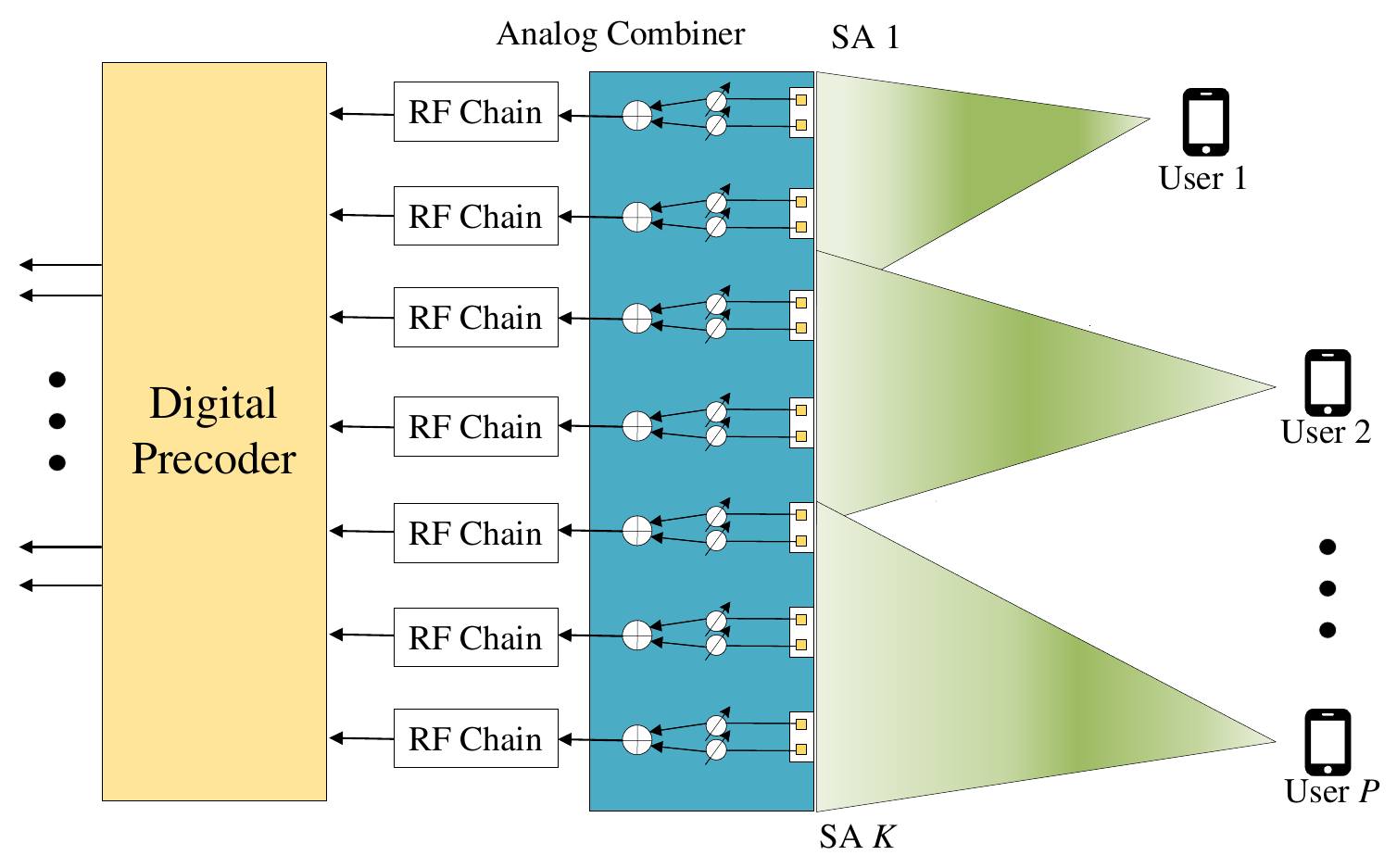}
	\caption{Hybrid beamforming structure of the modular XL-array.}
	\label{fig2}
\end{figure}

The sub-connected HBF architecture is applied to the modular XL-MIMO array, where the detailed structure is shown in Fig. 2. Specifically, each SA is connected to an RF chain via a phase shift network (PC-PSN), where each antenna within an SA is connected to a wideband analog phase-shifter.
Without loss of generality, we assume that the UEs simultaneously transmit pilot signals, and the received signal vector ${\bf{y}}\left[ i\right] \in {\mathbb{C}}^{K \times 1}$ of the whole XL-MIMO array at the $i$-th subcarrier can be expressed as
\begin{equation}\label{recsig}
\begin{split}
	{\bf{y}}\left [ i \right ] = & \left ( {\bf{F}}^{\rm{RF}}{\bf{F}}^{\rm{BB}}\left [  i\right ]  \right )^{\rm{H}}\left( {\bf{H}}\left [ i \right ]{\bf{s}} + {\bf{v}}\left[ i\right] \right)\\
	= & {\bf{F}}^{\rm{H}}\left [  i\right ]{\bf{H}}\left [ i \right ]{\bf{s}} + {\bf{F}}^{\rm{H}}\left [  i\right ]{\bf{v}}\left[ i\right],
\end{split}
\end{equation}
where ${\bf{s}} \in {\mathbb{C}}^{P\times 1}$ denotes the transmit pilot symbols of the $P$ UEs, ${\bf{H}}\left[ i\right] = [{\bf{h}}_1\left[ i\right],{\bf{h}}_2\left[ i\right],\cdots, {\bf{h}}_{P}\left[ i\right]] \in {\mathbb{C}}^{M \times P}$ denotes the XL-MIMO channel matrix of the $i$-th subcarrier, and ${\bf{v}}\left[ i\right] \sim \mathcal{CN}\left({\bf{0}}, {\sigma}^{2}{\bf{I}}_{M} \right) $ denotes the the complex-valued additive white Gaussian noise (AWGN), respectively.
In addition, ${\bf{F}}^{\rm{H}}\left[ i\right] $ denotes the hybrid combining matrix at the BS cascaded by ${\bf{F}}^{\rm{RF}} \in {\mathbb{C}}^{M \times K}$ and ${\bf{F}}^{\rm{BB}}\left[ i\right]  \in {\mathbb{C}}^{K \times K}$, which respectively denote the corresponding analog and digital combiner of the whole modular XL-MIMO array. Owing to the sub-connected architecture, the analog combining matrix ${\bf{F}}^{\rm{RF}}$ is a block diagonal matrix and can be expressed as
\begin{equation}\label{anacbn}
	\begin{split}
		{\bf{F}}^{\rm{RF}} = &\ {\rm{blkdiag}}\left \{ {\bf{f}}_{1}^{\rm{RF}},{\bf{f}}_{2}^{\rm{RF}} , \cdots, {\bf{f}}_{K}^{\rm{RF}}\right \}\\ 
		= &\begin{bmatrix}
			{\bf{f}}_{1}^{\rm{RF}} & {\bf{0}}_{M_{\rm{S}}\times1} & \cdots & {\bf{0}}_{M_{\rm{S}}\times1} \\
			{\bf{0}}_{M_{\rm{S}}\times1} & {\bf{f}}_{2}^{\rm{RF}} & \cdots &  {\bf{0}}_{M_{\rm{S}}\times1} \\
			\vdots & \vdots & \ddots & \vdots \\
			{\bf{0}}_{M_{\rm{S}}\times1} & {\bf{0}}_{M_{\rm{S}}\times1} & \cdots & {\bf{f}}_{K}^{\rm{RF}}
		\end{bmatrix},
	\end{split}
\end{equation}
where ${\bf{f}}^{\rm{RF}}_{k} \in {\mathbb{C}}^{M_{\rm{S}}\times 1}, k = 1, 2, \cdots, K$ denotes the analog combiner of the $k$-th SA.

\subsection{Geometric and Terahertz Channel Model}\label{chamod}
In order to obtain the balance between positioning accuracy and low complexity of the algorithm, we apply HSPWM in the modular XL-MIMO-based model, where the SWM is considered among SAs and PWM is considered within each SA, as depicted in Fig. 1. Specifically, the azimuth and elevation AoAs at the SAs are different from each other, while the antennas within an SA share common azimuth and elevation AoAs.

For further manipulations in terms of position estimation, we respectively denote the Cartesian coordinates of the $p$-th UE, the $l$-th scatterer, and the reference position of the $k$-th SA by ${\bf{q}}^{\rm{U}}_p = [ x^{\rm{U}}_p, y^{\rm{U}}_p, {z^{\rm{U}}_p}]^{\rm{T}}$, ${\bf{q}}_{l}^{\rm{C}} = [ x^{{\rm{C}}}_l, y^{{\rm{C}}}_l, z^{{\rm{C}}}_l]^{\rm{T}} $ and ${\bf{q}}_{k}^{\rm{B}} = [ x^{{\rm{B}}}_k, y^{{\rm{B}}}_k, z^{{\rm{B}}}_k]^{\rm{T}}$. In addition, the azimuth and elevation AoAs, and the distance from the $p$-th UE to the $k$-th SA are denoted as $\theta^{\rm{BU}}_{k,p}$, $\phi^{\rm{BU}}_{k,p}$ and $d^{\rm{BU}}_{k,p}$, respectively. Since all the SAs are deployed parallel to the $xOz$ plane, the AoAs and the distance can be further expressed as follows
\begin{subequations}\label{georel}
\begin{align}
	\theta^{\rm{BU}}_{k,p} &= {\rm{arctan}} \frac{x^{\rm{U}}_p-x^{{\rm{B}}}_k}{y^{\rm{U}}_p-y^{{\rm{B}}}_k},\\
	\phi^{\rm{BU}}_{k,p} &= {\rm{arctan}} \frac{z^{\rm{U}}_p-z^{{\rm{B}}}_k}{\left ( x^{\rm{U}}_p-x^{{\rm{B}}}_k \right ){\rm{sin}}\theta^{\rm{BU}}_{k,p}  +   \left ( y^{\rm{U}}_p-y^{{\rm{B}}}_k \right ){\rm{cos}}\theta^{\rm{BU}}_{k,p} },\\
	d^{\rm{BU}}_{k,p} &=\sqrt{\left ( x^{\rm{U}}_p-x^{{\rm{B}}}_k \right )^2+\left ( y^{\rm{U}}_p-y^{{\rm{B}}}_k \right )^2+\left ( z^{\rm{U}}_p-z^{{\rm{B}}}_k \right )^2}.
\end{align}	
\end{subequations}
In addition, the azimuth and elevation AoAs and distance from the $l$-th scatterer to the $k$-th SA are respectively denoted as $\theta^{\rm{BC}}_{k,l}$, $\phi^{\rm{BC}}_{k,l}$ and $d^{\rm{BC}}_{k,l}$, while the geometrical relationship can be similarly obtained.

In terms of the THz channel model, signals suffering multiple reflections and scatterings are severely attenuated owing to the path fading in the THz band. Therefore, we only consider the LoS channels and NLoS channels with single reflection between the UEs and the SAs, while the impact of potential reflections and scatterings in the propagation scenario is neglected.
In addition, the impact of propagation path loss and molecular absorption are both considered, which are both dependent on the carrier frequency and propagation distance.
Accordingly, the path loss of the LoS and NLoS path can be respectively modeled as \cite{han2015multiray}
\begin{equation}\label{losgan}
	{g}_{\rm{LoS}}\left ( f_i, d^{\rm{BU}}_{k,p} \right ) = \left ( \frac{c}{4\pi f_i d^{\rm{BU}}_{k,p}} \right )^{\frac{\alpha}{2}} {\rm{e}}^{-\frac{1}{2}{\mathcal{K}}\left ( f_i \right ) d^{\rm{BU}}_{k,p}},
\end{equation}
\begin{equation}\label{nlsgan}
	{g}_{\rm{NLoS}}\left ( f_i, d_{1}, d_{2} \right ) =  \left | \Gamma \right |{\rm{e}}^{j\vartheta}  {g}_{\rm{LoS}}\left ( f_i, d_{1} \right ){g}_{\rm{LoS}}\left ( f_i, d_{2} \right ),
\end{equation}
where $\alpha$ denotes the path loss exponent, and ${\mathcal{K}}\left( f_i \right) $ denotes the molecular absorption coefficient\cite{pan2022sumrate, boulogeorgos2018distance}, respectively. In terms of the NLoS channel, $\Gamma$ denotes the reflection coefficient as modeled in \cite{han2015multiray}, $\vartheta$ denotes the random phase shift, and $d_1$, $d_2$ denote the partitioned distances of the cascaded channel, respectively.

Then, since the BS is capable of conducting the signal processing on different sub-bands independently and simultaneously, we consider the transmission and processing on an arbitrary sub-band, and omit the index $i$ temporarily for the rest of this section.
By utilizing the PWM based on the far-field assumption, the LoS channel between the $p$-th UE and the $k$-th SA can be modeled as
\begin{equation}\label{loschn}
	{\bf{h}}^{\rm{LoS}}_{k,p} = {g}_{\rm{LoS}}\left (d^{\rm{BU}}_{k,p} \right ){\rm{e}}^{-j2\pi f \tau_{k,p}^{\rm{BU}}}{\bf{b}}_{M_{\rm{S}}}\left ( \theta^{\rm{BU}}_{k,p}, \phi^{\rm{BU}}_{k,p}\right ),
\end{equation}
and
\begin{equation}
	{\bf{b}}_{M_{\rm{S}}}\left ( \theta^{\rm{BU}}_{k,p}, \phi^{\rm{BU}}_{k,p}\right ) = {\bf{a}}_{M_{x}}\left ( \theta^{\rm{BU}}_{k,p}, \phi^{\rm{BU}}_{k,p}\right )\otimes{\bf{a}}_{M_{z}}\left (\phi^{\rm{BU}}_{k,p}\right ),
\end{equation}
where $\tau^{\rm{BU}}_{k,p} = d^{\rm{BU}}_{k,p}/c$ denotes the time of arrival (ToA) of the LoS link from the $p$-th UE to the $k$-th SA, and ${\bf{b}}_{M_{\rm{S}}}\left ( \theta^{\rm{BU}}_{k,p}, \phi^{\rm{BU}}_{k,p}\right )$ denotes the array steering vector w.r.t. the AoAs $\theta^{\rm{BU}}_{k,p}$ and ${\phi^{\rm{BU}}_{k,p}}$ at the $k$-th SA, respectively. In addition, we define the virtual AoAs as
\begin{equation}\label{ag2vag}
	\omega_{k,p}^{\rm{BU}} = {\rm{sin}}\theta_{k,p}^{\rm{BU}}{\rm{cos}}\phi_{k,p}^{\rm{BU}}, \varphi_{k,p}^{\rm{BU}} = -{\rm{sin}}\phi_{k,p}^{\rm{BU}},
\end{equation}
and the steering vector can be further denoted as
\begin{equation}
	{\bf{b}}_{M_{\rm{S}}}\left ( \omega_{k,p}^{\rm{BU}}, \varphi_{k,p}^{\rm{BU}}\right ) = {\bf{c}}_{M_{x}}\left ( \omega_{k,p}^{\rm{BU}} \right )\otimes{\bf{c}}_{M_{z}}\left (\varphi_{k,p}^{\rm{BU}}\right ),
\end{equation}
where
\begin{equation}
	{\bf{c}}_{G}\left(\varpi \right)  = [1, {\rm{e}}^{j{\frac{2\pi}{\lambda}}d{\rm{sin}}\varpi},\cdots, {\rm{e}}^{j{\frac{2\pi}{\lambda}}\left( G-1\right) d{\rm{sin}}\varpi}]^{\rm{H}} \in {\mathbb{C}}^{G\times1},
\end{equation}
where $G \in \{M_x, M_z\}$ and $\varpi \in \{\omega_{k,p}^{\rm{BU}}, \varphi_{k,p}^{\rm{BU}}\}$, respectively.
Similarly, the NLoS channel from the $p$-th UE via the $l$-th scatterer to the $k$-th SA is given by
\begin{equation}\label{nlschn}
	{\bf{h}}^{\rm{NLoS}}_{k,l,p} = {g}_{\rm{NLoS}}\left (d^{\rm{CU}}_{l,p},d^{\rm{BC}}_{k,l}  \right ){\rm{e}}^{-j2\pi f \tau_{k,l,p}^{\rm{NLoS}}}{\bf{b}}_{M_{\rm{S}}}\left ( \omega^{\rm{BC}}_{k,l}, \varphi^{\rm{BC}}_{k,l}\right ),
\end{equation}
where $d^{\rm{CU}}_{l,p}$ denotes the distance between the $p$-th UE and the $l$-th scatterer, and $\tau_{k,l,p}^{\rm{NLoS}} = \left(d_{l,p}^{\rm{CU}} + d_{k,l}^{\rm{BC}} \right)/c $ denotes the ToA of the corresponding NLoS path, respectively. Hence, the uplink channel vector from the $p$-th UE to the whole modular XL-MIMO array is represented by ${\bf{h}}_p = \left [{\bf{h}}^{\rm{H}}_{1,p}, {\bf{h}}^{\rm{H}}_{2,p}, \cdots, {\bf{h}}^{\rm{H}}_{K,p}  \right ]^{\rm{H}}$, and we have
\begin{equation}\label{saychn}
	{\bf{h}}_{k,p} = \chi_{k,0,p}{\bf{h}}_{k,p}^{\rm{LoS}} + \sum_{l=1}^{L} \chi_{k,l,p}{\bf{h}}_{k,l,p}^{\rm{NLoS}},
\end{equation}
which denotes the channel from the $p$-th UE to the $k$-th SA. It is worth noting that the scalar $\chi_{k,l,p} \in \{0, 1\}$ is applied to manifest the SNS property of the XL-MIMO array. In specific, $\chi_{k,l,p} = 0$ indicates that the $l$-th NLoS path between the $k$-th SA and the $p$-th UE does not exist, since the corresponding scatterer does not locate in the VR of the $k$-th SA. In addition, the value of  $\chi_{k,0,p}$ corresponds to the existence of the LoS path.

\begin{remark}
	The sub-connected HBF architecture presented herein provides multiple advantages for the proposed positioning system. By employing only one RF link per SA, we reduce the total number of RF chains from $M_{\rm{S}} \times K$ to $K$, significantly lowering hardware complexity and power consumption. The block diagonal structure of ${\bf{F}}^{\rm{RF}}$ enables us to decouple the signal processing across SAs while maintaining sufficient spatial resolution. Furthermore, despite the reduction in RF chains, our phase-shifting network design effectively captures multi-dimensional information across time, frequency, and spatial domains. This approach not only addresses the challenges of SNS in XL-MIMO systems, but also provides an efficient framework that balances positioning accuracy with implementation complexity in THz systems.
\end{remark}

Based on the aforementioned model, we propose a low-complexity 3-D positioning algorithm to obtain the localization of the UEs based on HSPWM, which is divided into three stages and sequentially introduced in the following three sections. Specifically, we first obtain the estimations of the AoAs w.r.t. typical visible SAs with the largest received power. Then, we derive the coarse Cartesian coordinate estimations of the UEs by utilizing the aforementioned AoA estimations based on the WLS algorithm. Finally, an RD-CS-based AoA estimation method with relatively low complexity is applied to refine the Cartesian coordinate estimation. The overall procedure is summarized in Fig. 3.
\begin{figure}[htbp]
	\centering
	\includegraphics[width=3.5in]{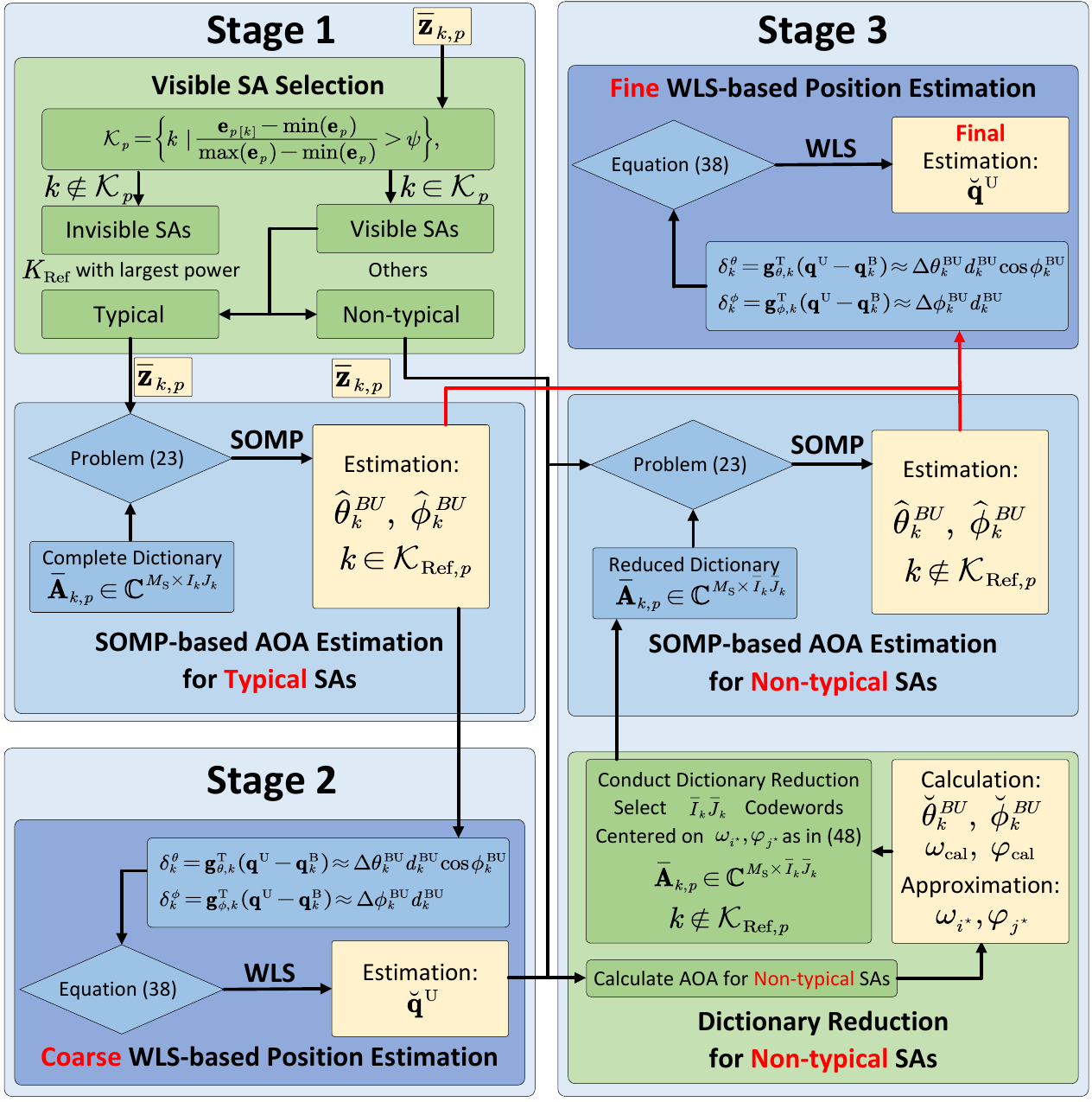}
	\caption{Flowchart of the localization framework.}
	\label{fig11}
\end{figure}

\section{Stage 1: Preliminary Processing \\ and AoA Estimation of Typical Visible SAs}\label{proal1}
In this section, we present a comprehensive framework for the initial stage of our positioning algorithm, which features several key innovations specifically designed for modular XL-MIMO systems. First, we elaborate on the training phase architecture, encompassing a specialized HBF configuration for our sub-connected architecture and orthogonal pilot sequence design. Then, we introduce a novel two-tier SA selection methodology to address the SNS problem in XL-MIMO systems: Firstly identifying visible SAs within the VR corresponding to each UE, followed by selecting typical visible SAs based on received signal power metrics. Finally, we implement a frequency-domain enhanced SOMP-based algorithm optimized for our HSPWM to obtain high-precision AoA estimations from these typical visible SAs, achieving abalance between positioning accuracy and complexity.

\subsection{Estimation Procedure and Preliminary Signal Processing}
Fig. 4 illustrates the specialized training stage design for our modular XL-MIMO architecture. In particular, the analog combiner of the modular XL-MIMO shifts for $N$ times during the whole training phase, which effectively compensates for the reduction in RF chains while maintaining sufficient spatial information capture. We term the time duration of fixed analog combiner as a ``training block'' consisting of $T$ time slots.

In the duration of each block, the $P$ UEs transmit orthogonal pilot sequences simultaneously, where the pilot symbols at the $t$-th time slot is denoted as 
${\bf{s}}_t = [s_{1,t}, s_{2,t}, \cdots, s_{P,t}]^{\rm{T}} \in {\mathbb{C}}^{P\times 1},$
and the pilot sequence of the $p$-th UE is denoted as
$\bar{\bf{s}}_p = [s_{p,1}, s_{p,2}, \cdots, s_{p,T}]^{\rm{T}} \in {\mathbb{C}}^{T\times 1},$
respectively. It is worth noting that one UE applies the same pilot sequence in all sub-carriers and in all the block durations, and we have ${\mathbb{E}}\left\lbrace\bar{\bf{s}}^{\rm{H}}_p\bar{\bf{s}}_p \right\rbrace = p_t$, where ${p_t}$ denotes the total transmit power of the users during $T$ time slots. Furthermore, the pseudo-random sequences such as PN sequences are utilized as the orthogonal pilot signals between different users, i.e.,  ${{\mathbb{E}}\left\lbrace\bar{\bf{s}}^{\rm{H}}_q\bar{\bf{s}}_p \right\rbrace = 0, q \ne p}$.

\begin{figure}[htbp]
	\centering
	\includegraphics[width=3in]{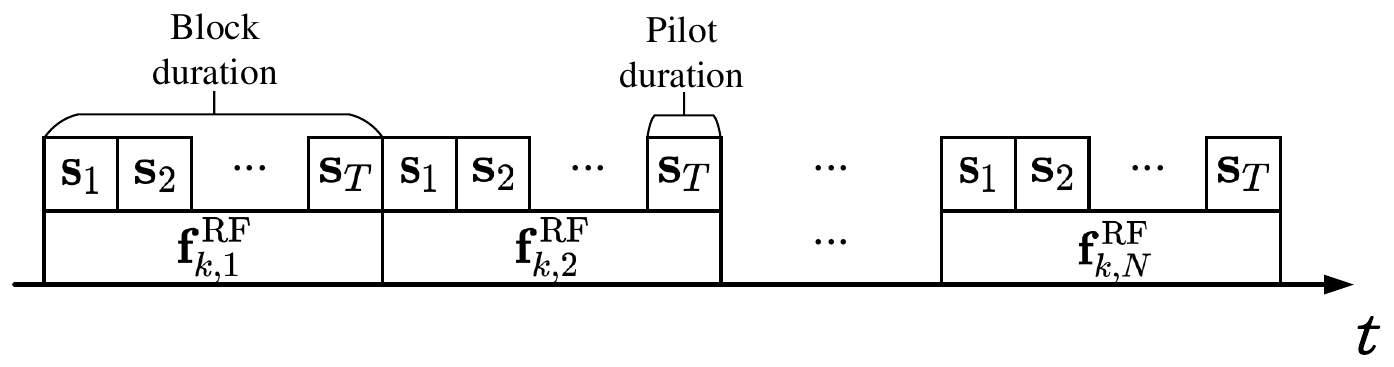}
	\caption{Training analog combiner and pilot design.}
	\label{fig3}
\end{figure}

We then introduce the HBF design during the training phase. For one thing, the digital combining matrix can be considered as ${\bf{F}}^{\rm{BB}} = {\bf{I}}_{K}$. Hence, the $k$-th entry of the receive signal ${\bf{y}}$ is actually the output of the analog combiner of the $k$-th SA, thereby decoupling the channel coefficients and the corresponding geometric parameters w.r.t. each SA. For another thing, the analog combiner of the SAs are independently and randomly generated in each block. Specifically, we have the $i$-th entry of ${{\bf{f}}^{\rm{RF}}_{k,n}}$ as ${\bf{f}}^{\rm{RF}}_{k,n[i]} ={\rm{e}}^{-j\psi_{k,n,i}}/\sqrt{M_{\rm{S}}}$, where ${\psi_{k,n,i}} =2\pi(n-1)(i-1)/N$.

Based on the aforementioned HBF and pilot design, we can derive  the output of the $k$-th SA in the $t$-th time slot of the $n$-th block as
\begin{equation}\label{atnsig}
	y_{k,n,t} = ({{\bf{f}}^{\rm{RF}}_{k,n}})^{\rm{H}}{\bf{H}}_{k}{\bf{s}}_{t} + ({{\bf{f}}^{\rm{RF}}_{k,n}})^{\rm{H}}{\bf{v}}_{k,n,t},
\end{equation}
where ${\bf{H}}_{k} = [{\bf{h}}_{k,1}, {\bf{h}}_{k,2}, \cdots, {\bf{h}}_{k,P}] \in {\mathbb{C}}^{M_{\rm{S}}\times P}$ denote the channels between the UEs and the $k$-th SA, and ${\bf{v}}_{k,n,t}$ denotes the corresponding thermal noise, respectively. 

Then, by stacking the received signal of the $k$-th SA during the $n$-th block in columns, we have
\begin{equation}\label{stkplt}
	\begin{split}
		{\bf{y}}_{k,n}^{\rm{T}} &= [y_{k,n,1}, y_{k,n,2},\cdots,y_{k,n,T}] \\
		&=({\bf{f}}^{\rm{RF}}_{k,n})^{\rm{H}}{\bf{H}}_{k}[{\bf{s}}_1,{\bf{s}}_2,\cdots,{\bf{s}}_{T}] + \bar{\bf{v}}_{k,n}^{\rm{T}}\\
		&=({\bf{f}}^{\rm{RF}}_{k,n})^{\rm{H}}{\bf{H}}_{k}{\bf{S}} + \bar{\bf{v}}_{k,n}^{\rm{T}},
	\end{split}
\end{equation}
where ${\bf{S}} = [{\bf{s}}_1,{\bf{s}}_2,\cdots,{\bf{s}}_{T}] = [\bar{\bf{s}}_1,\bar{\bf{s}}_2,\cdots,\bar{\bf{s}}_{P}]^{\rm{T}}\in{\mathbb{C}}^{{P\times T}}$, and ${\bar{\bf{v}}}_{k,n}$ denotes the corresponding thermal noise, respectively. In addition, we stack the vectors ${\bf{y}}_{k,n}^{\rm{T}}$ of the $N$ blocks in rows, and obtain the $N\times T$ receive matrix of the $k$-th SA during the whole training stage as follows
\begin{equation}\label{rcvmtx}
	\begin{split}
		{\bf{Y}}_{k} &= [{\bf{y}}_{k,1}, {\bf{y}}_{k,2},\cdots,{\bf{y}}_{k,N}]^{\rm{T}} \\
		&=(\bar{\bf{F}}^{\rm{RF}}_{k})^{\rm{H}}{\bf{H}}_{k}{\bf{S}} + \bar{\bf{V}}_{k}\\
		&=(\bar{\bf{F}}^{\rm{RF}}_{k})^{\rm{H}}\sum_{p=1}^{P} {\bf{h}}_{k,p}\bar{\bf{s}}_{p}^{\rm{T}} + {\bar{\bf{V}}}_{k},
	\end{split}
\end{equation}
where $\bar{\bf{F}}^{\rm{RF}}_{k} = [{\bf{f}}^{\rm{RF}}_{k,1}, {\bf{f}}^{\rm{RF}}_{k,2}, \cdots,{\bf{f}}^{\rm{RF}}_{k,N}]\in{\mathbb{C}}^{ M_{\rm{S}}\times N}$, and ${\bar{\bf{V}}}_{k} = [{\bar{\bf{v}}}_{k,1}, {\bar{\bf{v}}}_{k,2}, \cdots, {\bar{\bf{v}}}_{k,N}]^{\rm{T}}$, respectively. Then, by utilizing the orthogonal property of the pilot sequences, we can decouple the received vector corresponding to the $k$-th UE as follows
\begin{equation}\label{rcvvct}
	\begin{split}
		{\bf{z}}_{k,p} \triangleq {\bf{Y}}_{k}\bar{\bf{s}}_{p}^{\ast} & =  (\bar{\bf{F}}^{\rm{RF}}_{k})^{\rm{H}}\sum_{p=1}^{P} {\bf{h}}_{k,p}\bar{\bf{s}}_{p}^{\rm{T}}\bar{\bf{s}}_{p}^{\ast} + \bar{\bf{V}}_{k} \bar{\bf{s}}_{p}^{\ast}\\
		& = (\bar{\bf{F}}^{\rm{RF}}_{k})^{\rm{H}}{\bf{h}}_{k,p} p_t + \bar{\bf{n}}_{k,p}.
	\end{split}
\end{equation}

Based on these steps, the required channel parameters are decoupled in ${\bf{z}}_{k,p}$, which only contains the information of the channel between the $k$-th SA and the $p$-th UE.

\subsection{Selection of Visible SAs}
Unlike conventional approaches that assume all SAs can receive signals from all UEs, we explicitly model the SNS property of XL-MIMO arrays. Since the location of the scatters are also considered as unknown, the LoS paths towards all the SAs are essential to the positioning algorithm. Nevertheless, the LoS path between some UEs and some SAs may be obstructed owing to the SNS property of the proposed scenario.

Accordingly, we define the vector ${\bf{e}}_p = [\| {\bf{z}}_{1,p} \|_2, \| {\bf{z}}_{2,p} \|_2, \cdots, \| {\bf{z}}_{K,p} \|_2]^{\rm{T}}$ for each UE. Since we assume that the loss of NLoS paths is much higher than that of the LoS path, the received power at visible SAs should highly exceeds that of invisible ones. Hence, we obtain the set of visible SAs with LoS paths to the $p$-th UE with the following criterion
\begin{equation}\label{dtgctr}
	\mathcal{K}_p = \left \{k\ |\frac{{\bf{e}}_{p[k]}-\min({\bf{e}}_{p})}{\max({\bf{e}}_{p})-\min({\bf{e}}_{p})} >\psi \right \},
\end{equation}
where $0<\psi<1$ is a pre-defined threshold. By utilizing this normalized power-based criterion, we select the ``visible SAs'' w.r.t. each UE as localization anchors, while the others are considered as ``invisible''. If an SA is considered as invisible, the SA is totally eliminated in further AoA estimation and positioning algorithm w.r.t. the corresponding UE\footnote{Some of the SAs actually in the VR with relatively low power may also be excluded according to this criterion. This is acceptable since our purpose is selecting suitable SAs as positioning anchors instead of precisely determining the VR w.r.t. each UE.}.

Furthermore, we propose a progressive computational resource allocation strategy where the $K_{\rm{Ref}}$ SAs corresponding to the $K_{\rm{Ref}}$ largest values in ${\bf{e}}_p$ are selected as ``typical visible SAs'', which will be utilized for AoA and coarse position estimation in \textit{Stage 1} and \textit{2}, while other visible SAs, termed as ``non-typical visible SAs'', will be utilized for low-complexity fine estimation in \textit{Stage 3}. This two-tier selection mechanism significantly reduces computational complexity while maintaining positioning accuracy. For simplicity, the set of the indices of the typical visible SAs is denoted as ${\mathcal{K}}_{p,\rm{Ref}}$.

\subsection{SOMP-based AoA Estimation of Typical Visible SAs}
In terms of the proposed AoA estimation method, we leverage the unique properties of our HSPWM where each antenna within an SA shares the same azimuth and elevation AoAs to estimate the LoS channel between typical visible SAs and the UEs based on the SOMP method.

We first focus on the LoS channel parameter estimation between an arbitrary $k$-th SA and the $p$-th UE.
Based on HSPWM where each antenna within an SA shares the same azimuth and elevation AoAs, the angular-space representation of ${\bf{h}}_{k,p}$ can be denoted as
\begin{equation}\label{aglrps}
	{\bf{h}}_{k,p} = {\bf{A}}_{k,p}{\bm{\beta}}_{k,p},
\end{equation}
where ${\bf{A}}_{k,p} = [{\bf{b}}_{M_{\rm{S}}} ( \omega_{k,p}^{\rm{BU}}, \varphi_{k,p}^{\rm{BU}} ), {\bf{b}}_{M_{\rm{S}}} ( \omega_{k,1,p}^{\rm{BC}}, \varphi_{k,1,p}^{\rm{BC}} ), \cdots,$ $ {\bf{b}}_{M_{\rm{S}}} ( \omega_{k,L,p}^{\rm{BC}}, \varphi_{k,L,p}^{\rm{BC}} )] \in {\mathbb{C}}^{M_{\rm{S}} \times (L+1)}$ and ${{\bm{\beta}}_{k,p}}\in {\mathbb{C}}^{(L+1)\times 1}$ denotes the corresponding complex channel gain vector, respectively. Subsequently, according to the angular-space representation of the channels, we can formulate a CS problem to obtain the channel estimation as
\begin{equation}\label{cspbfm}
	\begin{split}
	{\bf{z}}_{k,p} & = p_t (\bar{\bf{F}}^{\rm{RF}}_{k})^{\rm{H}}\bar{\bf{A}}_{k,p}\bar{\bm{\beta}}_{k,p} + \bar{\bf{n}}_{k,p}\\
	& = p_t \bar{\bm{\Xi}}_{k,p}\bar{\bm{\beta}}_{k,p} + \bar{\bf{n}}_{k,p},
	\end{split}
\end{equation}
where the matrix $\bar{\bf{A}}_{k,p} = [{\bf{b}}_{M_{\rm{S}}}\left ( \omega_{1}, \varphi_{1}\right ), {\bf{b}}_{M_{\rm{S}}}\left ( \omega_{1}, \varphi_{2}\right ), \cdots,$ ${\bf{b}}_{M_{\rm{S}}}\left ( \omega_{I_k}, \varphi_{J_k}\right )]\in {\mathbb{C}}^{M_{\rm{S}} \times {I_k}{J_k}}$ denotes the angular-domain dictionary matrix of the $k$-th SA, and $\bar{\bm{\beta}}_k = [\bar\beta_{1,1}, \bar\beta_{1,2}, \cdots, \bar\beta_{I_k,J_k}]^{\rm{T}} \in {\mathbb{C}}^{{I_k}{J_k} \times 1}$ denotes the corresponding sparse vector with no more than $(L+1)$ non-zero elements, respectively. In specific, the $I_k$ samples of $\omega$ and  $J_k$ samples of $\varphi$ are uniformly sampled in $[-\pi, \pi]$, amounting to $I_kJ_k$ codewords in the dictionary. It can be derived that the entries of the sensing matrix $\bar{\bm{\Xi}}_{k,p} \triangleq {(\bar{\bf{F}}^{\rm{RF}}_{k})^{\rm{H}}\bar{\bf{A}}_{k,p}}$ satisfy the independent and identical distributed (i.i.d.) complex Gaussian distribution with zero mean and identical variance, indicating that the sensing matrix is suitable for sparse signal reconstruction.

Hence, the problem of (\ref{cspbfm}) can be solved via CS methods such as orthogonal matching pursuit (OMP) in each sub-band. For diversity in frequency, we can also utilize the SOMP method to address the problem on all sub-carriers by solving one CS problem owing to the common support property. 

To be specific, by concatenating the signals received at the $I$
sub-bands, we have
\begin{equation}\label{blkfml}
	\begin{split}
	{\bf{\bar{z}}}_{k,p} & = [{\bf{{z}}}_{k,p}^{\rm{T}}[1], {\bf{{z}}}_{k,p}^{\rm{T}}[2], \cdots, {\bf{{z}}}_{k,p}^{\rm{T}}[I]]^{\rm{T}}\\
	& = p_t \bar{\bm{\Theta}}_{k,p}\bar{\bm{\gamma}}_{k,p} + \bar{\bm{\omega}}_{k,p},
	\end{split}
\end{equation}
where 
\begin{equation}
	\bar{\bm{\Theta}}_{k,p} = {\rm{blkdiag}}\left\lbrace\bar{\bm{\Xi}}_{k,p}[1], \bar{\bm{\Xi}}_{k,p}[2], \cdots, \bar{\bm{\Xi}}_{k,p}[I] \right\rbrace,
\end{equation}
\begin{equation}
	\bar{\bm{\gamma}}_{k,p} = [\bar{\bm{\beta}}_{k,p}^{\rm{T}}[1], \bar{\bm{\beta}}_{k,p}^{\rm{T}}[2], \cdots, \bar{\bm{\beta}}_{k,p}^{\rm{T}}[I]]^{\rm{T}},
\end{equation}
and $\bar{\bm{\omega}}_{k,p}$ denotes the corresponding thermal noise, respectively, Then, we construct the vector ${{\breve{\bm{\gamma}}}}_{k,p} = [{\bm{\gamma}}_{k,p,1}^{\rm{T}}, {\bm{\gamma}}_{k,p,2}^{\rm{T}}, \cdots, {\bm{\gamma}}_{k,p,{I_k}{J_k}}^{\rm{T}}]^{\rm{T}}$, where ${{{\bm{\gamma}}_{k,p,j}}}$ is subsequently composed of the $j$-th elements of vectors $\bar{\bm{\beta}}_{k,p}[1], \bar{\bm{\beta}}_{k,p}[2], \cdots, \bar{\bm{\beta}}_{k,p}[I]$. It is obvious that $\left\lbrace \bar{\bm{\beta}}_{k,p}[i] \right\rbrace_{i=1}^{I}$ have common support, indicating that $\breve{\bm{\gamma}}_{k,p}$ is a block-sparse vector.

By denoting the corresponding permutation matrix as ${\bf{P}} \in \mathbb{C}^{I_kJ_kI \times I_kJ_kI}$, we reformulate (\ref{blkfml}) into a block sparse recovery problem as 
\begin{equation}\label{blkrfm}
	\begin{split}
		{\bf{\bar{z}}}_{k,p} & = p_t (\bar{\bm{\Theta}}_{k,p}{\bf{P}}^{\rm{T}})({\bf{P}}\bar{\bm{\gamma}}_{k,p}) + \bar{\bm{\omega}}_{k,p}\\
		& = p_t \breve{\bm{\Theta}}_{k,p}\breve{\bm{\gamma}}_{k,p} + \bar{\bm{\omega}}_{k,p},
	\end{split}
\end{equation}
where the permutation matrix is given by
\begin{equation}
	{\bf{P}}_{[u,v]} = \begin{cases}
		1, & {u = i + I(j-1) , v = j + I_kJ_k(i-1),} \\
		0, & {\textrm{else},}
	\end{cases}
\end{equation}
where $i = 1, 2, \cdots, I$, and $j = 1, 2, \cdots, I_kJ_k$, respectively. Then, we obtain the AoA estimation via the SOMP-based method as summarized in \textbf{Algorithm 1}. Note that we only focus on AoAs corresponding to the LoS path, we only need to select the block support index with the largest correlation owing to the much higher channel gain of the LoS path.
\begin{algorithm}
	\caption{SOMP-based AoA estimation for typical SAs.}
	\begin{algorithmic}[1]
		\State \textbf{Input:}  ${\bf{\bar{z}}}_{k,p},  I_k$, $J_k$, $k \in {\mathcal{K}}_{p,\rm{Ref}}$.
		\State Formulate the block sparse recovery problem as (\ref{blkrfm}).
    	\State Calculate the correlations as 
		\begin{equation}
			{\bf{d}} = \breve{\bm{\Theta}}_{k,p}^{\rm{H}}{\bf{\bar{z}}}_{k,p}.
	    \end{equation}
		\State Estimate the block support index as 
		\begin{equation}
			{\iota^{\star} = \left\lbrace \iota\ | \max_{1\le \iota \le {I_k}{J_k}}\sum_{s =(\iota-1)I+1}^{\iota I} |{\bf{d}}_{[s]}|\right\rbrace  },
		\end{equation}
		and the elements within the block as 
		\begin{equation}
			{\mathcal{J}}^{\star} = \left\lbrace (\iota^{\star}-1)I+1, (\iota^{\star}-1)I+2, \cdots, \iota^{\star}I  \right\rbrace.
		\end{equation}
		\State Obtain the virtual AoA estimations as ${\hat\omega^{\rm{BU}}_{k,p}} = \omega_{i^\star}$, and ${\hat\varphi^{\rm{BU}}_{k,p}} = \varphi_{j^\star}$, where $\iota^{\star} = j^\star + J_k(i^\star - 1)$, respectively.
		\State Calculate the AoA estimations ${\hat\theta^{\rm{BU}}_{k,p}}$ and ${\hat\phi^{\rm{BU}}_{k,p}}$ as (\ref{ag2vag}).
		\State \textbf{Output:}  ${\hat\theta^{\rm{BU}}_{k,p}}$ and ${\hat\phi^{\rm{BU}}_{k,p}}$.
	\end{algorithmic}
\end{algorithm}

\subsection{LoS Channel and ToA Estimation}
In addition to AoA estimation, accurate ToA information is capable of providing crucial distance measurements that significantly enhance the positioning accuracy in 3D space. Unlike traditional approaches that rely solely on AoA for positioning, our integrated framework incorporates the MUSIC algorithm for high-resolution ToA estimation, providing distance information which is also valuable for 3D localization tasks.

We can estimate the ToA based on the LoS channel estimations on all the sub-bands. Specifically, it is worth noting that the LoS channel gain on all sub-bands can be estimated as
\begin{equation}
	{\bf{g}}_{\iota^{\star}} = \left(\breve{\bm{\Theta}}_{k,p[:,{\mathcal{J}}^{\star}]}^{\rm{H}}\breve{\bm{\Theta}}_{k,p[:,{\mathcal{J}}^{\star}]}\right)^{-1}\breve{\bm{\Theta}}_{k,p[:,{\mathcal{J}}^{\star}]}^{\rm{H}}{\bf{\bar{z}}}_{k,p},
\end{equation}
where the $i$-th element of ${\bf{g}}_{\iota^{\star}}$ corresponds to the channel gain estimation on the $i$-th sub-band, i.e., $\hat{g}_{\rm{LoS}}\left (f_i, d^{\rm{BU}}_{k,p} \right ) = {\bf{g}}_{\iota^{\star}[i]}$.
In addition, the channel estimation w.r.t. the LoS channel between the $k$-th SA and the $p$-th UE on the $i$-th sub-band is given as
\begin{equation}
	\hat{\bf{h}}^{\rm{LoS}}_{k,p}[i]  = \hat{g}_{\rm{LoS}}\left (f_i, d^{\rm{BU}}_{k,p} \right ){\bf{b}}_{M_{\rm{S}}}\left ( \hat\omega^{\rm{BU}}_{k,p}, \hat\varphi^{\rm{BU}}_{k,p}\right ).
\end{equation}

It is worth noting that $\hat{g}_{\rm{LoS}}\left (f_i, d^{\rm{BU}}_{k,p} \right ) = |\Xi |{\rm{e}}^{j\omega}$, where the phase ${\omega}$ actually corresponds to $-2\pi f_i \tau_{k,p}^{\rm{BU}}$ in (\ref{loschn}). By leveraging the wideband nature of THz communications, we extract rich frequency-domain information for ToA estimation. We then substitute ${\bf{b}}_{M_{\rm{S}}}\left ( \hat\omega^{\rm{BU}}_{k,p}, \hat\varphi^{\rm{BU}}_{k,p}\right )$ with $\bar{\bf{b}}_{M_{\rm{S}}}\left ( \hat\omega^{\rm{BU}}_{k,p}, \hat\varphi^{\rm{BU}}_{k,p}\right )$ and construct another normalized vector as 
\begin{equation}
	\begin{split}
		\tilde{\bf{h}}^{\rm{LoS}}_{k,p}[i] & = {\rm{e}}^{-j2\pi f_i \tau_{k,p}^{\rm{BU}}}\bar{\bf{b}}_{M_{\rm{S}}}\left ( \hat\omega^{\rm{BU}}_{k,p}, \hat\varphi^{\rm{BU}}_{k,p}\right )\\
		& = {\rm{e}}^{-j2\pi (f_c+\psi[i]) \tau_{k,p}^{\rm{BU}}}\bar{\bf{b}}_{M_{\rm{S}}}\left ( \hat\omega^{\rm{BU}}_{k,p}, \hat\varphi^{\rm{BU}}_{k,p}\right ),
	\end{split}
\end{equation}
where $\psi[i] = f_i - f_c = \frac{B}{I}\left( i-\frac{I-1}{2}\right)$.

Then, by stacking $	\tilde{\bf{h}}^{\rm{LoS}}_{k,p}[i]$ in columns, we have
\begin{equation}
	\begin{split}
		\tilde{\bf{H}}_{k,p} & = [\tilde{\bf{h}}^{\rm{LoS}}_{k,p}[1], \tilde{\bf{h}}^{\rm{LoS}}_{k,p}[2], \cdots, \tilde{\bf{h}}^{\rm{LoS}}_{k,p}[I]]\\
		& = \bar{\bf{b}}_{M_{\rm{S}}}\left ( \hat\omega^{\rm{BU}}_{k,p}, \hat\varphi^{\rm{BU}}_{k,p}\right ) {\bf{e}}^{\rm{T}}(\tau_{k,p}^{\rm{BU}}),
	\end{split}
\end{equation}
where the vector ${\bf{e}}(\tau_{k,p}^{\rm{BU}}) = [{\rm{e}}^{-j2\pi (f_c+\psi[1]) \tau_{k,p}^{\rm{BU}}}, {\rm{e}}^{-j2\pi (f_c+\psi[2]) \tau_{k,p}^{\rm{BU}}}, \cdots, {\rm{e}}^{-j2\pi (f_c+\psi[I]) \tau_{k,p}^{\rm{BU}}}]^{\rm{T}}$.
We then conduct the EVD on the covariance matrix given as ${\bf{R}}_{\tilde{\bf{H}}_{k,p}^{\rm{H}}} = \tilde{\bf{H}}_{k,p}^{\rm{H}}\tilde{\bf{H}}_{k,p}$, obtain the eigenvectors corresponding to the noise subspace and stack them into a matrix as ${\bf{U}}_{\rm{z}}$. Subsequently, the ToA can be estimated as
\begin{equation}\label{mscsrc}
	\hat\tau_{k,p}^{\rm{BU}} = \left\lbrace \tau \ |  \max_{\tau \in (0,\tau_{\rm{max}})}\frac{1} {{\bf{e}}^{\rm{T}}(\tau){\bf{U}}_{\rm{z}}{\bf{U}}_{\rm{z}}^{\rm{H}}{\bf{e}}^{\ast}(\tau) }\right\rbrace,
\end{equation}
where $\tau_{\rm{max}}$ is a pre-defined upper-bound of $\tau$. This ToA information may serve as valuable complementary data which can be integrated in further manipulations to improve the overall positioning accuracy and robustness.

\section{Stage 2: Coarse Position Estimation}\label{proal2}
In this section, we introduce the second stage of our positioning framework, which transforms the estimated AoAs into Cartesian coordinates through a novel weighted least squares approach. Unlike conventional methods that directly use AoA estimates to approximate positions, our approach leverages geometric insights to formulate pseudo-linear equations that enable more accurate position estimation\footnote{For  simplicity in notations and better readability, we assume ${\mathcal{K}}_{p,\rm{Ref}} = \{1, 2, \cdots, K\}$ in the whole Section IV.}.

\subsection{The Construction of the Pseudo-linear Equations}
Specifically, we establish an effective relationship between AoA measurements and UE coordinates through carefully derived pseudo-linear equations. By utilizing the geometric relations as in (\ref{georel}a) and (\ref{georel}b), we can derive the equations in terms of the AoAs and the coordinates as follows
\begin{subequations}\label{pseqid}
	\begin{align}
		0 = &-\left ( x^{\rm{U}}-x^{\rm{B}}_k \right ) {\rm{cos}}{\theta^{\rm{BU}}_k}+\left ( y^{\rm{U}}-y^{\rm{B}}_k \right ) {\rm{sin}}{\theta^{\rm{BU}}_k},\\
		0 = &\left ( \left ( x^{\rm{U}}-x^{\rm{B}}_k \right ) {\rm{sin}}{\theta^{\rm{BU}}_k}+\left ( y^{\rm{U}}-y^{\rm{B}}_k \right ) {\rm{cos}}{\theta^{\rm{BU}}_k} \right ){\rm{sin}}{\phi^{\rm{BU}}_k}\nonumber \\
		& -\left ( z^{\rm{U}}-z^{\rm{B}}_k \right ) {\rm{cos}}{\phi^{\rm{BU}}_k}.
	\end{align}
\end{subequations}

Then, by substituting the realistic AoAs with the estimated AoAs, we have
\begin{subequations}\label{pseqrl}
	\begin{align}
		\delta^{\theta}_k = &-\left ( x^{\rm{U}}-x^{\rm{B}}_k \right ) {\rm{cos}}{\hat\theta^{\rm{BU}}_k}+\left ( y^{\rm{U}}-y^{\rm{B}}_k \right ) {\rm{sin}}{\hat\theta^{\rm{BU}}_k}  ,\\
		\delta^{\phi}_k = &\left ( \left ( x^{\rm{U}}-x^{\rm{B}}_k \right ) {\rm{sin}}{\hat\theta^{\rm{BU}}_k}+\left ( y^{\rm{U}}-y^{\rm{B}}_k \right ) {\rm{cos}}{\hat\theta^{\rm{BU}}_k} \right ){\rm{sin}}{\hat\phi^{\rm{BU}}_k}\nonumber \\
		& -\left ( z^{\rm{U}}-z^{\rm{B}}_k \right ) {\rm{cos}}{\hat\phi^{\rm{BU}}_k},
	\end{align}
\end{subequations}
where $\delta^{\theta}_k$ and $\delta^{\phi}_k$ denote the residual errors corresponding to ({\ref{pseqid}}a) and ({\ref{pseqid}}b) owing to the estimation errors of ${\theta}_k^{{\rm{BU}}}$ and ${\phi}_k^{{\rm{BU}}}$, respectively.
We next denote the estimation error as $\Delta\theta^{\rm{BU}}_k =\hat\theta^{\rm{BU}}_k -\theta^{\rm{BU}}_k$ and $\Delta\phi^{\rm{BU}}_k =\hat\phi^{\rm{BU}}_k -\phi^{\rm{BU}}_k$, which are assumed to be small satisfying ${\rm{sin}}\Delta\theta^{\rm{BU}}_k\approx \Delta\theta^{\rm{BU}}_k$, ${\rm{sin}}\Delta\phi^{\rm{BU}}_k\approx \Delta\phi^{\rm{BU}}_k$, ${\rm{cos}}\Delta\theta^{\rm{BU}}_k\approx 1$, and ${\rm{cos}}\Delta\phi^{\rm{BU}}_k\approx 1$, respectively. By substituting these approximations into (\ref{pseqrl}a) and (\ref{pseqrl}b), we can derive the approximation of the residual errors as follows
\begin{subequations}\label{aardap}
	\begin{align}
		\delta^{\theta}_k \approx & -\left ( x^{\rm{U}}-x^{\rm{B}}_k \right ) \left ( {\rm{cos}}{\theta^{\rm{BU}}_k}-\Delta\theta^{\rm{BU}}_k{\rm{sin}}{\theta^{\rm{BU}}_k} \right )\nonumber \\
		&+\left ( y^{\rm{U}}-y^{\rm{B}}_k \right ) \left ( {\rm{sin}}{\theta^{\rm{BU}}_k} + {\Delta\theta^{\rm{BU}}_k}{\rm{cos}}{\theta^{\rm{BU}}_k}\right )\nonumber\\
		=&\Delta\theta^{\rm{BU}}_k\left ( \left ( x^{\rm{U}}-x^{\rm{B}}_k \right ){\sin}{\theta^{\rm{BU}}_k}+\left ( y^{\rm{U}}-y^{\rm{B}}_k \right ){\cos}{\theta^{\rm{BU}}_k} \right )\nonumber \\
		=&\Delta\theta^{\rm{BU}}_k d_{k}^{\rm{BU}}{\cos}{{\phi}_{k}^{\rm{BU}}},\\
		\delta^{\phi}_k \approx &\ d^{\rm{BU}}_k {\cos}\phi_{k}^{\rm{BU}}\left ( {\rm{sin}}{\phi^{\rm{BU}}_k} + {\Delta\phi^{\rm{BU}}_k}{\rm{cos}}{\phi^{\rm{BU}}_k} \right )\nonumber \\
		&-d^{\rm{BU}}_k {\sin}\phi_{k}^{\rm{BU}} \left ( {\rm{cos}}{\phi^{\rm{BU}}_k}-\Delta\phi^{\rm{BU}}_k{\rm{sin}}{\phi^{\rm{BU}}_k} \right )\nonumber\\
		=&\ \Delta\phi^{\rm{BU}}_k d_k^{\rm{BU}}.
	\end{align}
\end{subequations}

For further manipulations, we define the direction vectors
\begin{equation}
	\begin{split}
		{\bf{g}}_{\theta,k} & = [ -{\rm{cos}}{\hat\theta^{\rm{BU}}_k}, {\rm{sin}}{\hat\theta^{\rm{BU}}_k}, 0 ]^{\rm{T}},\\
		{\bf{g}}_{\phi,k} & =  [ {\rm{sin}}{\hat\theta^{\rm{BU}}_k}{\rm{sin}}{\hat\phi^{\rm{BU}}_k}, {\rm{cos}}{\hat\theta^{\rm{BU}}_k}{\rm{sin}}{\hat\phi^{\rm{BU}}_k}, -{\rm{cos}}{\hat\phi^{\rm{BU}}_k} ]^{\rm{T}}.
	\end{split}
\end{equation}

Subsequently, we can obtain the pseudo-linear equations corresponding to the $k$-th SA as follows
\begin{subequations}\label{agpseq}
	\begin{align}
		\delta_k^{\theta} = {\bf{g}}_{\theta,k}^{\rm{T}} \left ({\bf{q}}^{\rm{U}} - {\bf{q}}_k^{\rm{B}}  \right ) & \approx \Delta\theta^{\rm{BU}}_k d_{k}^{\rm{BU}}{\cos}{{\phi}_{k}^{\rm{BU}}},\\
		\delta_k^{\phi} = {\bf{g}}_{\phi,k}^{\rm{T}} \left ({\bf{q}}^{\rm{U}} - {\bf{q}}_k^{\rm{B}}  \right ) & \approx \Delta\phi^{\rm{BU}}_k d_k^{\rm{BU}}.
	\end{align}
\end{subequations}

For further manipulations, we then derive the compact form of the aforementioned pseudo-linear equations. 
By combining Equations (\ref{agpseq}a) and (\ref{agpseq}b), we can obtain the following compact form of the pseudo-linear equations as
\begin{equation}\label{cptequ}
	\tilde{\bf{G}} {\bf{q}}^{\rm{U}} - \tilde{\bf{h}} = \tilde{\bf{D}}{\bf{z}},
\end{equation}
where
\begin{equation}\label{notcpt}
	\begin{split}
		\tilde{\bf{G}} = \left[	\tilde{\bf{G}}_{\theta}^{\rm{T}},	\tilde{\bf{G}}_{\phi}^{\rm{T}} \right]^{\rm{T}}, \tilde{\bf{h}} = \left[	\tilde{\bf{h}}_{\theta}^{\rm{T}}, \tilde{\bf{h}}_{\phi}^{\rm{T}} \right]^{\rm{T}},\\ 
		\tilde{\bf{D}} = \left[\tilde{\bf{D}}_{\theta}, \tilde{\bf{D}}_{\phi}\right]^{\rm{T}}, {\bf{z}} = \left[ {\bf{n}}_{\theta}^{\rm{T}}, {\bf{n}}_{\phi}^{\rm{T}}\right]^{\rm{T}}.
	\end{split}
\end{equation}

In terms of the expression of ${\tilde{\bf{G}}}$, we have
\begin{equation}\label{notggg}
	\begin{split}
		\tilde{\bf{G}}_{\theta} &= \left [ \tilde{\bf{g}}_{\theta,1}, \tilde{\bf{g}}_{\theta,2}, \cdots, \tilde{\bf{g}}_{\theta,K} \right ]^{\rm{T}} \in {\mathbb{C}}^{K \times 3},\\
		\tilde{\bf{G}}_{\phi} &= \left [ \tilde{\bf{g}}_{\phi,1}, \tilde{\bf{g}}_{\phi,2}, \cdots, \tilde{\bf{g}}_{\phi,K} \right ]^{\rm{T}}\in {\mathbb{C}}^{K \times 3}.
	\end{split}
\end{equation}

Then, in terms of the expression of ${\tilde{\bf{h}}}$, we have
\begin{equation}\label{nothhh}
	\begin{split}
		\tilde{\bf{h}}_{\theta} &= \left [ {\bf{g}}_{\theta,1}^{\rm{T}}{\bf{q}}_{1}^{\rm{B}}, {\bf{g}}_{\theta,2}^{\rm{T}}{\bf{q}}_{2}^{\rm{B}},\cdots,{\bf{g}}_{\theta,K}^{\rm{T}}{\bf{q}}_{K}^{\rm{B}}\right ]^{\rm{T}} \\
		&=\left ( \tilde{\bf{G}}_{\theta} \odot \tilde{\bf{Q}}_{\rm{B}}^{\rm{T}} \right ){\bf{1}}_{3\times1} \in {\mathbb{C}}^{K\times1}, \\
		\tilde{\bf{h}}_{\phi} &= \left [ {\bf{g}}_{\phi,1}^{\rm{T}}{\bf{q}}_{1}^{\rm{B}}, {\bf{g}}_{\phi,2}^{\rm{T}}{\bf{q}}_{2}^{\rm{B}},\cdots,{\bf{g}}_{\phi,K}^{\rm{T}}{\bf{q}}_{K}^{\rm{B}}\right ]^{\rm{T}} \\
		&=\left ( \tilde{\bf{G}}_{\phi} \odot \tilde{\bf{Q}}_{\rm{B}}^{\rm{T}} \right ){\bf{1}}_{3\times1} \in {\mathbb{C}}^{K\times1},
	\end{split}
\end{equation}
where $\tilde{\bf{Q}}_{\rm{B}} = \left[ {\bf{q}}^{\rm{B}}_1, {\bf{q}}^{\rm{B}}_2, \cdots, {\bf{q}}^{\rm{B}}_K\right]$.

Next, in terms of the expression of $\tilde{{\bf{D}}}$, we have $\tilde{\bf{D}}_{\theta} = [{\bf{D}}_{\theta}, {\bf{O}}_{K\times K}]^{\rm{T}}$, and $\tilde{\bf{D}}_{\phi} = [{\bf{O}}_{K\times K}, {\bf{D}}_{\phi}]^{\rm{T}}$, where
\begin{align}\label{notddd}
	{\bf{D}}_{\theta} &= {\rm{diag}}\left \{ d_1^{\rm{BU}}\cos\phi_1^{\rm{BU}}, d_2^{\rm{BU}}\cos\phi_2^{\rm{BU}}, \cdots, d_K^{\rm{BU}}\cos\phi_K^{\rm{BU}}\right \},\nonumber\\
	{\bf{D}}_{\phi} &= {\rm{diag}}\left \{ d_1^{\rm{BU}}, d_2^{\rm{BU}}, \cdots, d_K^{\rm{BU}}\right \}.
\end{align}

Finally, in terms of the expression of ${{\bf{z}}}$, we have
\begin{align}\label{notnnn}
{\bf{n}}_{\theta} &= [\Delta \theta_{1}^{\rm{BU}}, \Delta \theta_{2}^{\rm{BU}}, \cdots, \Delta \theta_{K}^{\rm{BU}} ]^{\rm{T}},\nonumber\\
{\bf{n}}_{\phi} &= [\Delta \phi_{1}^{\rm{BU}}, \Delta \phi_{2}^{\rm{BU}}, \cdots, \Delta \phi_{K}^{\rm{BU}} ]^{\rm{T}},
\end{align}
respectively.

\subsection{The Procedure of the WLS Positioning Algorithm}
Our iterative weighted least squares approach offers a significant advantage over traditional closed-form solutions by optimally accounting for the varying reliability of measurements from different SAs. This is particularly important in XL-MIMO systems where measurement quality can vary substantially across sub-arrays due to spatial non-stationarity.

Based on the aforementioned compact form of pseudo-linear equations, we can derive the position estimation of the UE by applying the iterative WLS algorithm as follows.

According to the key idea of the WLS method, we firstly construct the WLS cost function of (\ref{cptequ}) as
\begin{equation}\label{cosfun}
	C\left ( \tilde{\bf{q}}^{\rm{U}} \right )  =  ( \tilde{\bf{G}} \tilde{\bf{q}}^{\rm{U}} - \tilde{\bf{h}}  )^{\rm{T}}{\bf{W}} ( \tilde{\bf{G}} \tilde{\bf{q}}^{\rm{U}} - \tilde{\bf{h}}  ),
\end{equation}
where ${\bf{W}}$ denotes the weight matrix of the WLS method, which is constructed as 
\begin{equation}\label{wgtmtx}
	{\bf{W}} = {\bf{R}}_{\rm{e}}^{-1}, {\bf{R}}_{\rm{e}} = {\mathbb{E}}\left \{ \tilde{\bf{D}}{\bf{z}}{\bf{z}}^{\rm{T}} \tilde{\bf{D}}^{\rm{T}}\right \}  = \tilde{\bf{D}}{\bf{R}}_{\rm{z}} \tilde{\bf{D}}^{\rm{T}},
\end{equation}
where ${\bf{R}}_{\rm{z}}$ is the covariance of the estimation error ${\bf{z}}$. Since the estimation error of each parameter is independent of each other, ${\bf{R}}_{\rm{z}}$ is a block diagonal matrix as ${\bf{R}}_{\rm{z}}  =  {\rm{blkdiag}}\left \{ {\bf{R}}_{\Delta\theta},{\bf{R}}_{\Delta\phi}\right \}$.

Then, by letting the first-order derivative of the WLS cost function in (\ref{cosfun}) equal to 0, i.e., $\partial C ( \tilde{\bf{q}}^{\rm{U}}  ) /\partial \tilde{\bf{q}}^{\rm{U}} = \bf{0}$, we can minimize the cost function and obtain the closed-form estimation of $\tilde{\bf{q}}^{\rm{U}}$. Specifically, the first-order derivative of $C ( \tilde{\bf{q}}^{\rm{U}}  )$ is given by
\begin{equation}\label{devequ}
	\frac{\partial C ( \tilde{\bf{q}}^{\rm{U}}  )}{\partial \tilde{\bf{q}}^{\rm{U}}  }  = 2\tilde{\bf{G}}^{\rm{T}}{\bf{W}}\tilde{\bf{G}}\tilde{\bf{q}}^{\rm{U}}-2\tilde{\bf{G}}^{\rm{T}}{\bf{W}}\tilde{\bf{h}}.
\end{equation}
Hence, the coarse position estimation $\breve{\bf{q}}^{\rm{U}} = [\breve{x}^{\rm{U}}, \breve{y}^{\rm{U}}, \breve{z}^{\rm{U}}]^{\rm{T}}$ can be derived as
\begin{equation}\label{coaest}
	\breve{\bf{q}}^{\rm{U}} =   ( \tilde{\bf{G}}^{\rm{T}}{\bf{W}}\tilde{\bf{G}}  ) ^{-1}\tilde{\bf{G}}^{\rm{T}}{\bf{W}}\tilde{\bf{h}}.
\end{equation}

 Specifically, we initialize ${\bf{W}}$ as ${\bf{I}}$ and obtain the initial estimation of $\breve{\bf{q}}^{\rm{U}}$, then we update the value of ${\bf{W}}$ based on the initial position estimation and repeat this process iteratively. This progressive refinement ensures convergence to a highly accurate solution even with imperfect AoA measurements. The detailed procedure of the iterative WLS method is summarized in \textbf{Algorithm 2}.

\begin{algorithm}
	\caption{The iterative WLS procedure.}
	\begin{algorithmic}[1]
		\State \textbf{Input:} ${\hat\theta^{\rm{BU}}_{k,p}}$ and ${\hat\phi^{\rm{BU}}_{k,p}}$.
		\State Initialize the WLS cost matrix ${\bf{W}} = {\bf{I}}$.
		\State {\textbf{Repeat:}}
		\State \quad Update the estimation $\breve{\bf{q}}^{\rm{U}}$ via (\ref{coaest}).
		\State \quad Update the distances ${d}^{\rm{BU}}_k$ via (\ref{georel}) and $\breve{\bf{q}}^{\rm{U}}$.
    	\State \quad Update the $\tilde{\bf{D}}$ and ${\bf{W}}$ via (\ref{notddd}), (\ref{wgtmtx}) and ${d}^{\rm{BU}}_k$.
		\State {\textbf{Until}} {\textit{Convergence}}.
		\State \textbf{Output:}  $\breve{\bf{q}}^{\rm{U}}$.
	\end{algorithmic}
\end{algorithm}

\section{Stage 3: Low Complexity \\Fine Position Estimation}\label{proal3}
In this final stage, we present a computationally efficient approach to refine the position estimate by leveraging all visible SAs while maintaining low complexity. Our innovative reduced-dictionary SOMP method represents a significant advancement over conventional approaches that either sacrifice accuracy for reduced complexity or require prohibitively high computational resources for high accuracy.

\subsection{RD-based AoA Estimation of Non-typical Visible SAs}
Our RD-based AoA estimation approach is capable of dramatically reducing the computational complexity without sacrificing estimation accuracy. Although the AoAs of the LoS paths of the non-typical visible SAs can also be similarly obtained via the SOMP-based AoA estimation as described in \textit{Stage 1}, we further aim to reduce the complexity via dictionary reduction based on geometric relations.

In particular, after obtaining the coarse position estimations of the UEs in \textit{Stage 2}, we can calculate the AoAs w.r.t. the non-typical visible SAs and take the calculations as a reference of estimation. This geometric-guided dictionary reduction strategy represents a significant departure from conventional approaches that employ fixed dictionaries regardless of prior information. 

\subsection{WLS-based Fine Position Estimation}
Based on \textbf{Algorithm 3}, the AoAs of all the visible SAs w.r.t. the same UE are estimated sequentially. Then, in terms of the $p$-th UE, we apply the iterative WLS method in \textbf{Algorithm 2} to obtain a fine position estimation, which utilizes full AoA information w.r.t. all the visible SAs. In specific, angular information of selected typical and non-typical visible SAs should be both utilized in this process. This hierarchical estimation approach, proceeding from coarse to fine, enables our framework to achieve both computational efficiency and high accuracy simultaneously.

\begin{algorithm}
	\caption{RD-SOMP-based AoA Estimation of Non-typical Visible SAs.}
	\begin{algorithmic}[1]
		\State \textbf{Input:} Coarse position estimation $\breve{\bf{q}}^{\rm{U}}_p$, non-typical SA receive vector $\bar{\bf{z}}_{k,p}$, $k \notin \mathcal{K}_{p,\rm{Ref}}$, and $\bar i$, $\bar j$.
		\State Calculate the reference AoAs w.r.t. the $k$-th SA and the $p$-th UE as $\breve\theta_{k,p}^{\rm{BU}}$ and $\breve\phi_{k,p}^{\rm{BU}}$ via $\breve{\bf{q}}^{\rm{U}}$,  (\ref{georel}a) and (\ref{georel}b).
		\State Calculate the virtual AoAs corresponding to $\breve\theta_{k,p}^{\rm{BU}}$ and $\breve\phi_{k,p}^{\rm{BU}}$ via (\ref{ag2vag}) as $\omega_{\rm{cal}}$, $\varphi_{\rm{cal}}$.
		\State Approximate $\omega_{\rm{cal}}$, $\varphi_{\rm{cal}}$ as $\omega_{i^{\star}}$, $\varphi_{j^{\star}}$ as the best on-grid approximation w.r.t. the original dictionary $\bar{\bf{A}}_{k,p}$. 
		\State Construct the corresponding reduced dictionary as
		\begin{equation}
			\begin{split}
				\tilde{\bf{A}}_{k,p} = [{\bf{b}}_{M_{\rm{S}}}\left ( \omega_{i^{\star}-{\bar{i}}}, \varphi_{j^{\star}-{\bar{j}}}\right ), \cdots, {\bf{b}}_{M_{\rm{S}}}\left ( \omega_{i^{\star}+{\bar{i}}}, \varphi_{j^{\star}-{\bar{j}}}\right ),\\ {\bf{b}}_{M_{\rm{S}}}\left ( \omega_{i^{\star}-{\bar{i}}}, \varphi_{j^{\star}-{\bar{j}}+1}\right ), \cdots, {\bf{b}}_{M_{\rm{S}}}\left ( \omega_{i^{\star}+{\bar{i}}}, \varphi_{j^{\star}+{\bar{j}}}\right )],
			\end{split}
		\end{equation}
		where $\tilde{\bf{A}}_{k,p} \in {\mathbb{C}}^{M_{\rm{S}}\times (2\bar{i}+1)(2\bar{j}+1)}$, and ${\bar{i}}$, ${\bar{j}}$ are pre-defined parameters, respectively. 
		\State Obtain the AoA estimations with the reduced dictionary $\tilde{\bf{A}}_{k,p}$ via \textbf{Algorithm 1}.
		\State \textbf{Output:} ${\hat\theta^{\rm{BU}}_{k,p}}$ and ${\hat\phi^{\rm{BU}}_{k,p}}$.
	\end{algorithmic}
\end{algorithm}
\subsection{Complexity Analysis}
  The computational complexity of the proposed framework is composed of the complexity of three stages. Specifically, the complexity of \textit{Stage 1} is dominated by the formulation of the sensing matrix and the process of SOMP, which amounts to ${\mathcal{O}}( K_{\rm{Ref}}I_kJ_kM_{\rm{S}}NI)$. The complexity of \textit{Stage 2} is dominated by the inverse operation to update $\bf{W}$, which is given by ${\mathcal{O}}({K^3_{\rm{Ref}}})$ in each iteration. Similarly, the complexity of \textit{Stage 3} is given as ${\mathcal{O}}( (K-K_{\rm{Ref}})\bar{I}_k\bar{J}_kM_{\rm{S}}NI + K^3) $, where $\bar{I}_k = 2\bar{i} + 1$ and $\bar{J}_k = 2\bar{j} + 1$, respectively. 

It is particularly noteworthy that our reduced dictionary approach decreases the sensing matrix size in \textit{Stage 3} by a quadratic order, thereby significantly reducing the corresponding complexity in this stage. Hence, it can be concluded that the overall computational complexity is dominated by $\mathcal{O}{(K_{\rm{Ref}}I_kJ_kM_{\rm{S}}NI)}$, which represents a substantial improvement over conventional methods requiring full dictionary processing for all SAs.

\section{Simulation Results}\label{simres}
In this section, simulation results are provided to illustrate the advantage of the proposed localization algorithm. The geometric setup of the simulation is depicted in Fig. 1.

\subsection{Simulation Setup and Benchmark Schemes}\label{simset}
\begin{table}[htbp]
	\renewcommand\arraystretch{2}
	\centering
	\caption{Default Simulation Parameters}
	\label{table1}
	\begin{tabular}{|c|c|}
		\hline
		\makecell[l]{System\\ configuration} & \makecell[l]{\linespread{2}\selectfont $B = $ 4 GHz, $N_{\rm{c}}$ = 1025, $f_{\rm{c}} =$ 320 GHz, \\ $I = $ 5, $P =$ 2, $L = $2,\\ $K = K_x \times K_z = $ 5 $ \times $ 5 ,\\ $M_{\rm{S}} = M_x \times M_z = $ 5 $\times $ 5,} \\ \hline
		\makecell[l]{Geometric \\ model} & \makecell[l]{\linespread{2}\selectfont ${\bf{q}}_1^{\rm{B}}$ = (0, 0, 0), $D = 1$ m, $d = \lambda_c$/4,\\ ${\bf{q}}_1^{\rm{C}}$ = (5m,5m,5m), ${\bf{q}}_2^{\rm{C}}$ = (-20m,5m,15m), \\
		${\bf{q}}_1^{\rm{U_o}}$ = (-3m,3m,1.5m), ${\bf{q}}_2^{\rm{U_o}}$ = (-5m,5m,2m),} \\ \hline 
		\makecell[l]{Transmission\\ model} & \makecell[l]{$\sigma^2 =$ -120 dBm/Hz, $T = $ 5, $N = $ 25, $\alpha$ = 2, } \\ \hline
		\makecell[l]{Signal \\processing} & \makecell[l]{$\Delta \omega $= 0.01, $\Delta \varphi $= 0.01, $\bar{i}$ = 8, $\bar{j}$ = 8.}\\ \hline
	\end{tabular}
\end{table}

Unless otherwise stated, the other simulation parameters are listed in Table 1, where ${\bf{q}}_1^{\rm{U_o}}$ and ${\bf{q}}_2^{\rm{U_o}}$ denote the center of the distributed area of UE1 and UE2.
The coefficient of the molecular absorption is modeled according to the simplified model for the 200 - 400 GHz band as in \cite{boulogeorgos2018distance}.

To evaluate the performance in 3-D positioning of the algorithms, we consider the root mean square error (RMSE) of the coordinate estimation as the metric.

Furthermore, in order to illustrate the advantage of the proposed HSPWM-based algorithm in complexity and accuracy, we also compare the proposed scheme with the benchmark schemes as follows. The complexity and running time comparison of the proposed scheme and benchmarks is shown in Table II, where the running time is corresponding to one sample in Fig. 5 and 6.
\begin{itemize}
	\item[1)]
	\textbf{Benchmark 1:}  In this scenario, the key idea of the joint channel estimation and positioning algorithm proposed in \cite{qiao2024sensing} is utilized. Some adjustments are applied to adapt to the differences between the system models: All UEs are considered as active UEs, and we substitute the 2-D positioning algorithm in \cite{qiao2024sensing} with the proposed WLS-based positioning method.
\end{itemize}

\begin{itemize}
	\item[2)]
	\textbf{Benchmark 2 (DFT-MUSIC in collocated design):} In this scenario, we consider the positioning scheme based on one $M_x \times M_z = $ 25 $\times$ 25 collocated array without HBF structure located at ${\bf{q}}_{1}^{\rm{B}}$. The 2D-DFT algorithm is applied to estimate the AoAs, and the sub-band-based method in Section III. D. is applied to estimate the ToA. The coordinate estimations are directly obtained via the estimated AoAs and ToA. The parameters $G_{\rm{x}}$ and $G_{\rm{z}}$ in Table II denote the search grids on the $x$-axis and $z$-axis of DFT, respectively.
\end{itemize}

\begin{table}[htbp]
	\renewcommand\arraystretch{2}
	\centering
	\caption{Complexity Comparison}
	\label{table2}
	\begin{tabular}{|c|c|c|}
		\hline
		\makecell[l]{Method} & \makecell[l]{Computational \\ Complexity} & \makecell[l]{Running Time  }  \\ \hline
		\makecell[l]{SOMP-WLS \\ with RD} & \makecell[l]{$\mathcal{O}{(K_{\rm{Ref}}I_kJ_kM_{\rm{S}}NI)}$} &\makecell[l]{12.2886s $(K_{\rm{Ref}} = 2)$\\17.8827s $(K_{\rm{Ref}} = 3)$}  \\ \hline 
		\makecell[l]{SOMP-WLS \\ without RD} & \makecell[l]{$\mathcal{O}{(KI_kJ_kM_{\rm{S}}NI)}$} &\makecell[l]{120.4797s}  \\ \hline 
		\makecell[l]{Benchmark 1} & \makecell[l]{$\mathcal{O}(KM^3_{\rm{S}}I)$} &\makecell[l]{32.4783s}  \\ \hline 
		\makecell[l]{Benchmark 2} & \makecell[l]{$\mathcal{O}(G_{\rm{x}}G_{\rm{z}}MI)$} &\makecell[l]{3.3654s}  \\ \hline 
	\end{tabular}
\end{table}
%

%

\subsection{Performance Comparison of the Proposed Algorithm}\label{percom}
Fig. 5 illustrates the positioning performance of the proposed framework with different $K_{\rm{Ref}}$ and the performance difference between the coarse and fine estimations. Note that in the scenario ``$K_{\rm{Ref}} = $ 25'', we apply the SOMP-based AoA estimation without RD to all the detected visible SAs. It can be regarded as an upper bound of the proposed framework, and the gap between it and other curves reveals the performance loss introduced by fewer utilized SAs or RD.

As depicted in Fig. 5 and Table II, the performance of the fine estimation is evidently superior to the coarse estimation, and close to the upper bound with lower computational complexity. For one thing, it is observed that better positioning performance can be achieved with more visible SAs utilized. Hence, the performance of fine estimation is expected to obviously exceed that of the coarse estimation with more directly estimated AoA information, even if the corresponding dictionaries are edited according to the coarse coordinate estimation. For another, since the coarse coordinate estimation is actually close to the real position to a certain degree, proper dictionary reduction will result in little extra AoA estimation error. Therefore, properly designed reduced dictionary is capable of reducing the complexity with ideal positioning performance.

Fig. 6 illustrates the comparison between the proposed framework and aforementioned benchmarks. As depicted in Fig. 6, the proposed scheme consistently outperforms all the other benchmarks, especially in low SINR scenarios.
Specifically, the positioning performance of the proposed modular XL-MIMO-based framework evidently exceeds that of the traditional 2-D DFT relying on one 25 $\times$ 25 collocated array.
It indicates that the modular-based design is capable of utilizing more angular information w.r.t. different SAs, thereby achieving better positioning performance compared with traditional collocated design.

In addition, the RMSE of the proposed framework is close to that of the method in \cite{qiao2024sensing} in high SINR scenario, while the performance gap becomes quite high with relatively low SINR. This phenomenon can be attributed to the fact that the AoA estimation is highly dependent on the preliminary LoS channel estimation via LS algorithm, which is highly sensitive to noise. Furthermore, it is worth noting that the DFT-MUSIC-based method also suffers severe degradation in low SINR scenario, which is owing to the error of in LoS channel gain introduced by the LS algorithm. 

These results obviously manifest the superiority of the proposed modular XL-MIMO design in localization, and the advantage of the proposed SOMP-WLS-based positioning algorithm within this framework. The proposed wireless localization framework efficiently utilizes the diversity in the frequency and spatial domain, which is robust to noise and capable of providing more accurate and reliable 3-D location estimation with lower complexity.
\begin{figure}[htbp]
	\centering
	\includegraphics[width=2.4in]{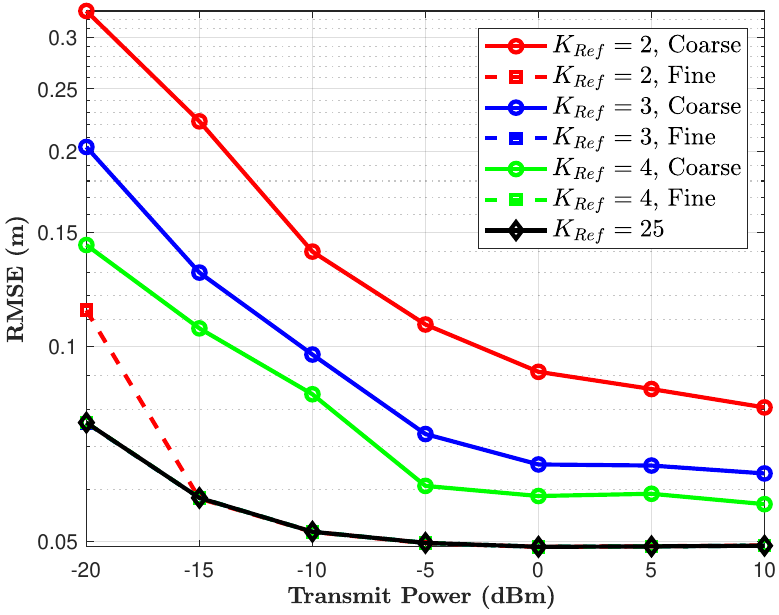}
	\caption{RMSE comparison  with different $K_{\rm{Ref}}$.}
	\label{fig4}
\end{figure}
\begin{figure}[htbp]
	\centering
	\includegraphics[width=2.4in]{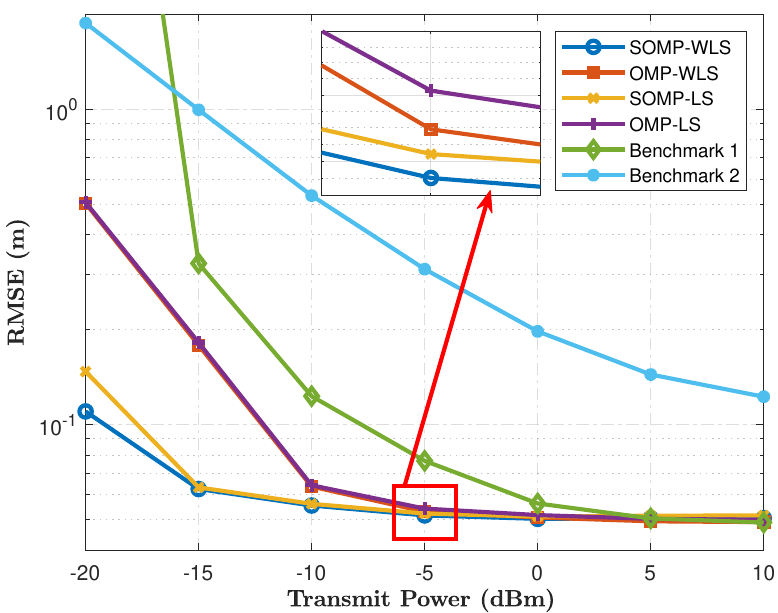}
	\caption{RMSE comparison  with different algorithms.}
	\label{fig5}
\end{figure}

\subsection{Impact of Interval between SAs}\label{ipctsa}
Fig. 7 illustrates the performance with different intervals between SAs. In particular, the RMSE  firstly descends and then ascends with the increase of SA interval, indicting that there exists an optimal SA interval for the least RMSE performance.

For one thing, the RMSE performance improvement with the increase of the SA interval can be attributed to better utilization of the spatial DoFs. Specifically, the disparity among the AoAs w.r.t. the same UE becomes larger, thereby providing more efficient dimensions to the WLS algorithm for better positioning performance. In contrast, the WLS algorithm may fail to obtain a reasonable position estimation owing to the low rank of matrix $\tilde{\bf{G}}$ when the SA interval becomes smaller ($D = $ 0.2 m in default simulation setup), which also verifies that AoA information will become insufficient for 3-D positioning  with low SA interval.

For the other, the decline in RMSE performance along with further increase in SA interval can be attributed to poor performance in AoA estimation, especially for the SAs with larger distance to the UE. Specifically, when the SA interval is set as 2 m, the RMSE corresponding to the position estimation derived via all the visible SAs is even worse than that of two typical visible SAs. This phenomenon indicates that since larger SA interval will result in longer distance and severer fading w.r.t. some of the visible SAs, the AoA estimations actually have negative impact on the final position estimation. Hence, the interval between SAs requires  specialized design for optimal positioning performance, where the distance between the SAs and UEs should be carefully considered.
\begin{figure}[htbp]
	\centering
	\includegraphics[width=2.4in]{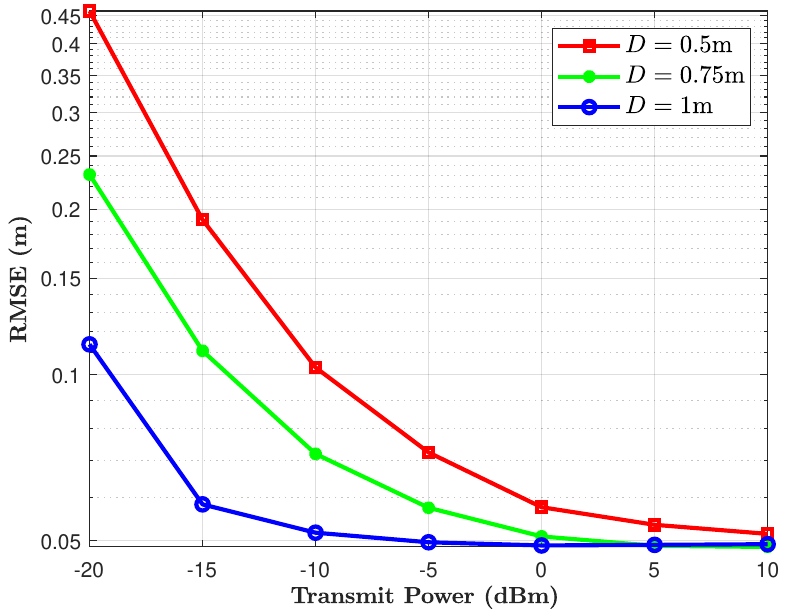}
	\label{fig8sub1}
	\\
	\includegraphics[width=2.4in]{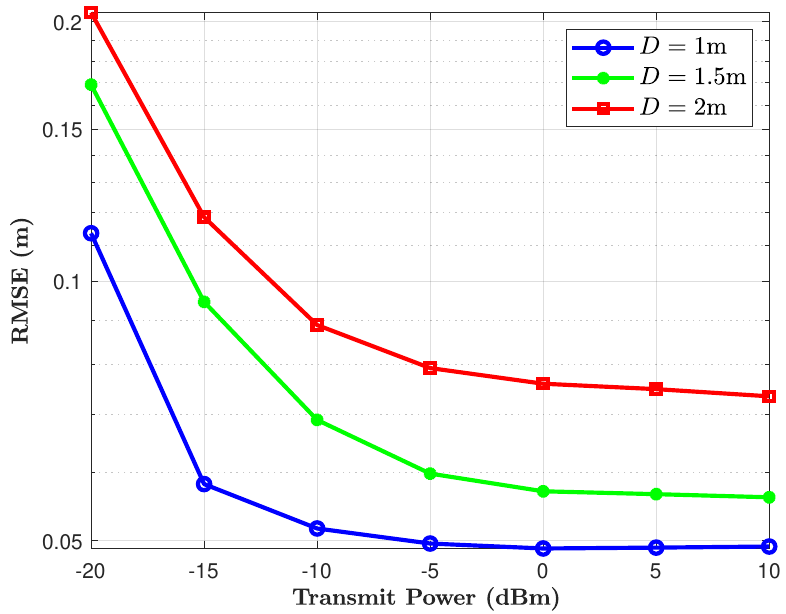}
	\label{fig8sub2}
	\caption{RMSE  with different SA interval.}
	\label{fig8}
\end{figure}

\subsection{Impact of AE Allocation}\label{ipctae}
Fig. 8 illustrates the impact of different allocation schemes with the same total number of the AEs. In this simulation setup, the total amount of the AEs is set as $M = $ 24 $\times$ 24 $=$ 576 with different allocation schemes, while we still set $M_x = M_z$ and $K_x = K_z$ for fairness in accuracy between the  azimuth and elevation orientation. 

It can be observed from Fig. 8 that along with the increase in the number of SAs, the RMSE descends at first and then ascends, reaching preferable performance when $K_x = K_z = $ 4 and $M_x = M_z = $ 6. This indicates that there exists a trade-off between the number of SAs and the number of AEs within each SA. In particular, the number of SAs represents the efficient angular information or spatial DoFs of the whole modular array, while the number of AEs within each SA represents the angular resolution corresponding to each AoA estimation. Such phenomenon reveals that there is a trade-off between the utilized spatial DoFs within the whole propagation region and the accuracy w.r.t. each AoA estimations. In addition, it is essential to properly design AE allocation to strike the balance between the two aspects, and thus obtain optimal positioning performance.
\begin{figure}[htbp]
	\centering
	\includegraphics[width=2.4in]{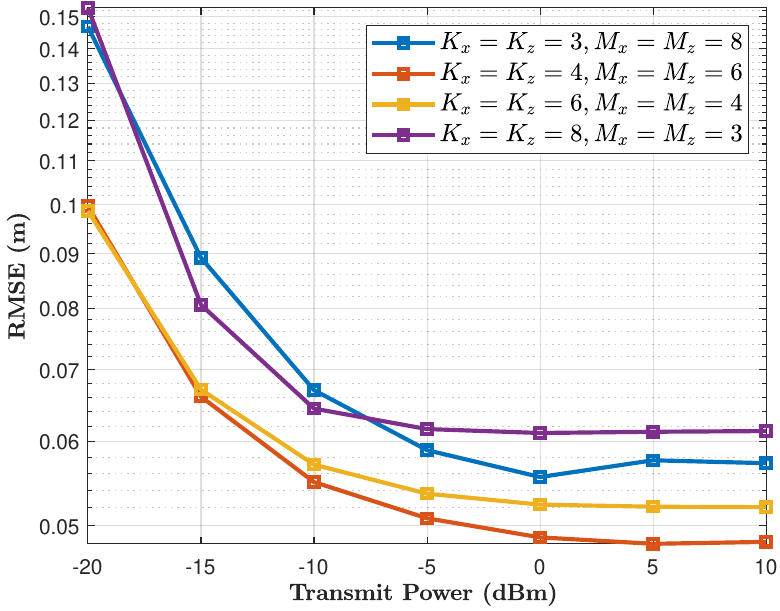}
	\caption{RMSE with different AE allocation.}
	\label{fig7}
\end{figure}

\subsection{Impact of Number of Training Blocks}\label{ipcttb}
Fig. 9 illustrates the relationship between the number of training blocks and the RMSE performance. It is worth noting that since the HBF structure combines the received signal of $M_{\rm{S}}$ AEs in one SA to one RF-chain, the number of training blocks $N$ should not be less than $M_{\rm{S}}$ to recover the dimension of channel between each UE and SA. As shown in Fig. 8, the positioning performance is improved with the increase in training blocks number, which is consistent to the expectation that more pilot symbols is able to support better performance in channel estimation and localization.

Nevertheless, the RMSE only reduced by 38.89$\%$ when $N$ increases from 25 to 50 with $P_t$ = -20 dBm, which corresponds to double pilot consumption. In contrast, the reduction in RMSE is 48.01$\%$ when the transmit power increases from -20 dBm to -15 dBm. In addition, the performance gap obtained by increasing block number becomes even smaller in high SINR scenario. Hence, it is reasonable to conclude that increasing the number of training blocks is not a sensible scheme to improve the positioning performance.
\begin{figure}[tbp]
	\centering
	\includegraphics[width=2.4in]{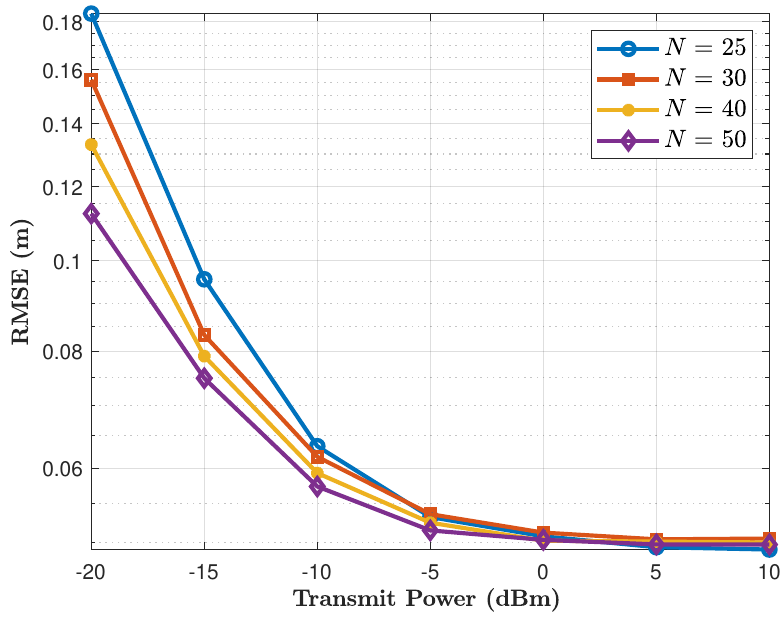}
	\caption{RMSE  with different training blocks numbers.}
	\label{fig9}
\end{figure}

\subsection{Impact of VR}\label{ipctvr}
Fig. 10 illustrates robustness of the proposed framework to the existence of VR. It is worth noting that  ``\textbf{VR = Diagonals}'' refers to the scheme where the 9 SAs on the diagonals of the whole 5 $\times$ 5 modular array are assumed to locate in the VR, and ``\textbf{VR =  3 $\times$ 3}'' refers to the scheme where an arbitrary group of 3 $\times$ 3 SAs are assumed to locate in the VR, respectively.
As depicted in Fig. 10, when only 9 out of 25 SAs are located in the VR,  the proposed framework is capable of selecting these visible SAs, obtaining corresponding AoA estimations, and deriving the final 3-D position estimation. 
Hence, we can conclude that the proposed framework is directly adaptive to the scenario with the existence of SNS in the modular XL-MIMO.

In addition, it can be observed that positioning performance of the schemes considering the existence of VR exhibits relative improvement along with the increase of transmit power at the BS.
Specifically, the ratio of the RMSE of ``\textbf{VR =  3 $\times$ 3}'' to ``\textbf{VR =  5 $\times$ 5}'' is 3.27 when $P_t = $ -20 dBm, which declines to 1.07 when $P_t = $ 10 dBm. 
This phenomenon can be attributed to the propagation of the error in AoA estimations. In particular, in high SINR scenarios, all SAs can provide AoA estimations with relatively low error, and 9 accurate AoA estimations are enough to support precise positioning with a slight gap to the scheme with 25 visible SAs.
In contrast, the positioning algorithm will suffer much severer error propagation when all SAs in the VR are relatively far to the UE with low SINR, while the performance will not be evidently improved when all SAs in the VR are relatively near to the UE owing to limited spatial DoFs.
Hence, when the VR is arbitrarily distributed in the range of the modular XL-array, the performance of the proposed algorithm with 9 visible SAs is acceptable in low SINR and close to the scheme without SNS in high SINR.

Furthermore, it is illustrated that the positioning performance of ``\textbf{VR = Diagonals}'' scheme slightly exceeds that of the ``\textbf{VR =  3 $\times$ 3}'' scheme. This can be attributed to the fact that the SAs deployed as diagonals is capable of providing more AoA information compared with SAs deployed as a square in the considered scenario, which is consistent to the speculation in subsection C.
\begin{figure}[tbp]
	\centering
	\includegraphics[width=2.4in]{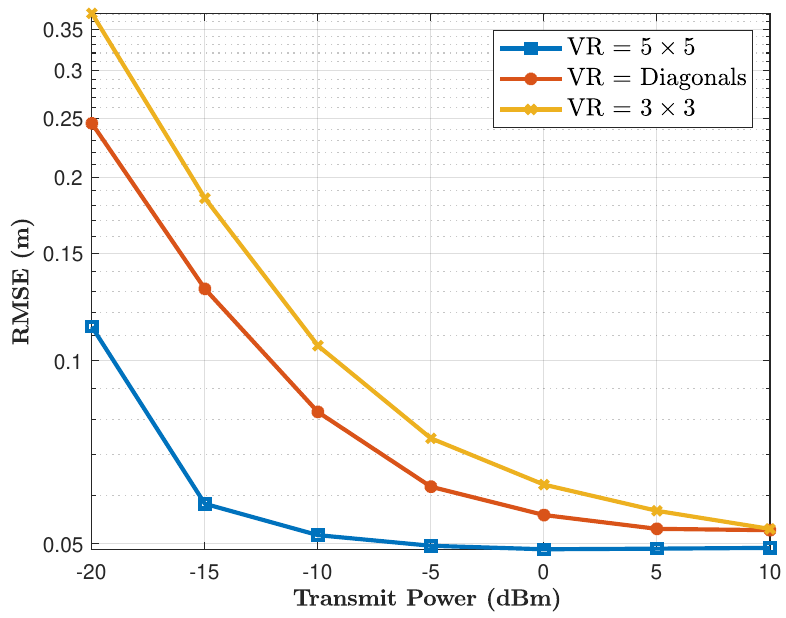}
	\caption{RMSE with different layout of the VR.}
	\label{fig6}
\end{figure}

\section{Conclusions}\label{conclu}
In this paper, we aimed at obtaining the Cartesian coordinate estimation of the UEs in a modular XL-array-enabled THz system with high accuracy and low complexity.
To handle this problem, we proposed a 3-D positioning framework based on HSPWM channel model and a three-stage localization algorithm.
In particular, we first introduced a training phase design for the modular architecture with HBF structure and the method to distinguish visible SAs located in the VR of each UE.
Subsequently, an SOMP-based algorithm was applied to estimate the AoAs w.r.t. typical visible SAs in \textit{Stage 1}.
Then, a closed-form coarse location estimation was iteratively derived based on the AoA information via WLS algorithm in \textit{Stage 2}.
Finally, we proposed an RD-CS-based AoA estimation method to estimate the AoAs of the rest visible SAs to reduce computational complexity, and refine the position estimation according to all efficient angle information in \textit{Stage 3}.
Simulation results indicated that the modular array-enabled positioning framework is capable of providing more accurate location estimation, and is more adaptive to scenarios with the existence of SNS or low SINR compared with its traditional centralized counterpart.
Furthermore, essential system factors such as the intervals between SAs and AE allocation should be carefully designed for better positioning performance of the proposed framework.


\bibliographystyle{IEEEtran}
\bibliography{20240920}

\end{document}